\documentclass{aa}

\usepackage{graphicx}
\usepackage{caption}
\usepackage{subcaption}
\usepackage{multirow}

\usepackage{txfonts}
\usepackage{xcolor}
\usepackage[breaklinks,colorlinks,urlcolor=blue,citecolor=blue,linkcolor=blue]{hyperref}

\begin{document}

   \title{Investigating white-light flare mechanisms via the Paschen jump using high-resolution continuum observations from the Swedish 1-m Solar Telescope}

   \author{S. Ornig\inst{1} \fnmsep \inst{2}\and
          L. Rouppe van der Voort\inst{1} \fnmsep \inst{2}\and
          M. Carlsson\inst{1} \fnmsep \inst{2}\and
          C. J. Díaz Baso\inst{1} \fnmsep \inst{2}\and
          E. S. Øyre\inst{1} \fnmsep \inst{2}\and
          I. J. Soler Poquet\inst{1}\fnmsep\inst{2}\and
          A. R. Brunvoll\inst{1} \fnmsep \inst{2}}

   \institute{Institute of Theoretical Astrophysics, University of Oslo,
              PO Box 1029 Blindern, 0315 Oslo, Norway\\
              \email{sascha.ornig@astro.uio.no}
         \and
             Rosseland Centre for Solar Physics, University of Oslo,
              PO Box 1029 Blindern, 0315 Oslo, Norway\\
             }

   \date{Received 14 April 2026 / Accepted 3 July 2026}
 
  \abstract 
   {The continuum is understood to contain a large portion of the energy emitted by a solar flare. The optical continuum, known as white light (WL), is particularly relevant since it may be observed by ground-based instruments.}
   {We measured the WL enhancements short- and longward of the Paschen jump in order to gain insights into the possible mechanism(s) behind the creation of these increases in our two case studies.}
   {We took measurements from the Swedish 1-m Solar Telescope of the pseudo-continuum around the \ion{Ca}{II}~8542~\AA~line as well as the true continuum around the \ion{K}{I}~7699~\AA~and the \ion{Fe}{I}~6173~\AA~line, providing us with observations on both sides of the Paschen jump.}
   {We observe WL enhancements of over 40\% against the dark (pen-) umbral background in both flares. No WL excess is detectable against the granulation outside the sunspots. The WL excess in flare~1 is co-temporal with the derivative of the GOES soft X-ray and hard X-ray (HXR) measurements from the Advanced Space-based Solar Observatory (ASO-S), and the flare is compatible with the Neupert effect. For the second flare, a preceding smaller flare may be the cause of the temporal discrepancy. Signatures of chromospheric evaporation and condensation are found in the WL area for both flares. The ratio of intensities blueward and redward of the Paschen jump (i.e., the Paschen ratio) in flare~1 is below one for most WL pixels. This is in disagreement with the accepted WL formation mechanisms, which are both of photospheric and chromospheric origin. We believe this is a consequence of the \ion{Ca}{II}~8542~\AA~pseudo-continuum being affected by line wing opacity changes.}
   {The co-temporality of WL and HXR enhancements suggests that the WL emission enhancements in flare~1 (and parts of flare~2) are a result of direct electron precipitation. We conclude that more reliable continuum measurements free of any nearby line influence are necessary in order to obtain conclusive evidence for the formation mechanism(s) behind optical continuum enhancements from such analysis as presented in this work.}


   \titlerunning{White-light flare mechanisms inferred using the Swedish 1-m Solar Telescope}
   \maketitle


\section{Introduction}

Solar flares are spatially and temporally localized brightenings on the Sun visible over the whole electromagnetic spectrum, lasting minutes to hours. In solar physics, the main focus over the last decades has been on spectral lines, which carry information highly relevant to many open questions. The continuum, on the other hand, generally suffers from poor contrast (due to the background intensity of the photosphere) and a short duration that events yield. However, the optical continuum --- white light (WL) --- can be observed with ground-based instruments at a high temporal and spatial resolution, offering a unique diagnostic window.\\
\indent The optical continuum carries a significant amount of the total flare energy. According to \cite{1989SoPh..121..261N}, WL intensity enhancements might contain as much as 90\% of the total energy output from a flare region. This hypothesis was reinforced by observations of the Sun as a star by \cite{2011A&A...530A..84K}, which revealed that nearly two-thirds of a flare's radiated energy is found in WL.\\
\indent Despite the plethora of observations from the ground, the mechanisms behind WL flares are still hotly debated. The theories can primarily be categorized into two groups. One posits that the emissions originate from the photosphere, where emissions at similar wavelengths are emitted in the quiet Sun. In this scenario, energy must be transported to the photosphere either by high-energy electron beams \citep{1978ApJ...224..241E, 2018ApJ...862...76P}, Alfvén waves \citep{2013ApJ...765...81R}, or through radiative backwarming \citep{1989SoPh..124..303M, 2003ApJ...595..483M}. Radiative backwarming involves heated layers in the solar atmosphere releasing radiation that warms the lower layers, leading to an increase in temperature and therefore intensity. The other category of theories suggests that the emission comes from higher regions, specifically the chromosphere. Chromospheric WL is identified by optically thin recombination radiation, predominantly in the hydrogen continua \citep{2018ApJ...867L..24D}.\\
\begin{figure*}
    \resizebox{\hsize}{!}
    {\includegraphics[width=0.33\textwidth]{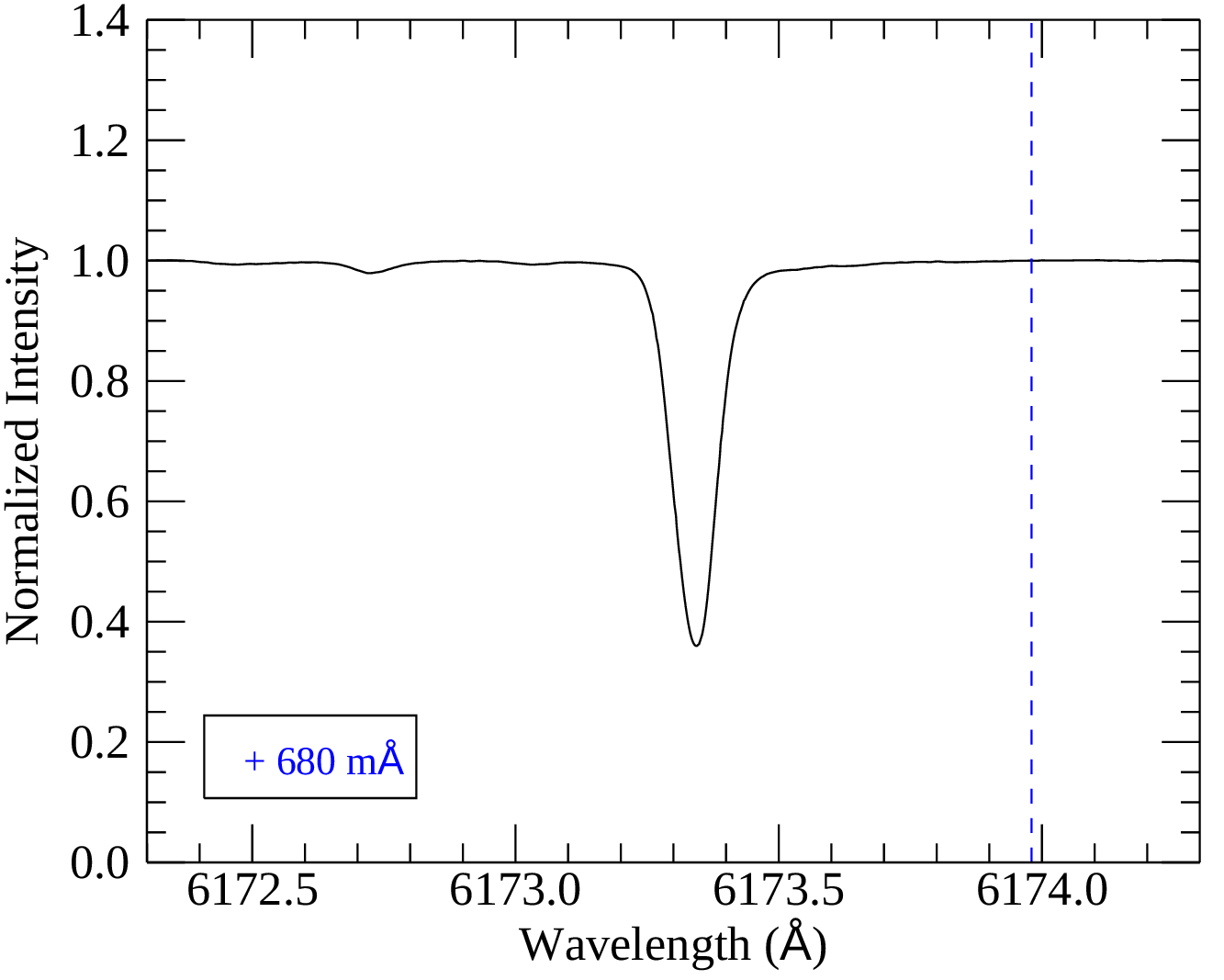}\includegraphics[width=0.33\textwidth]{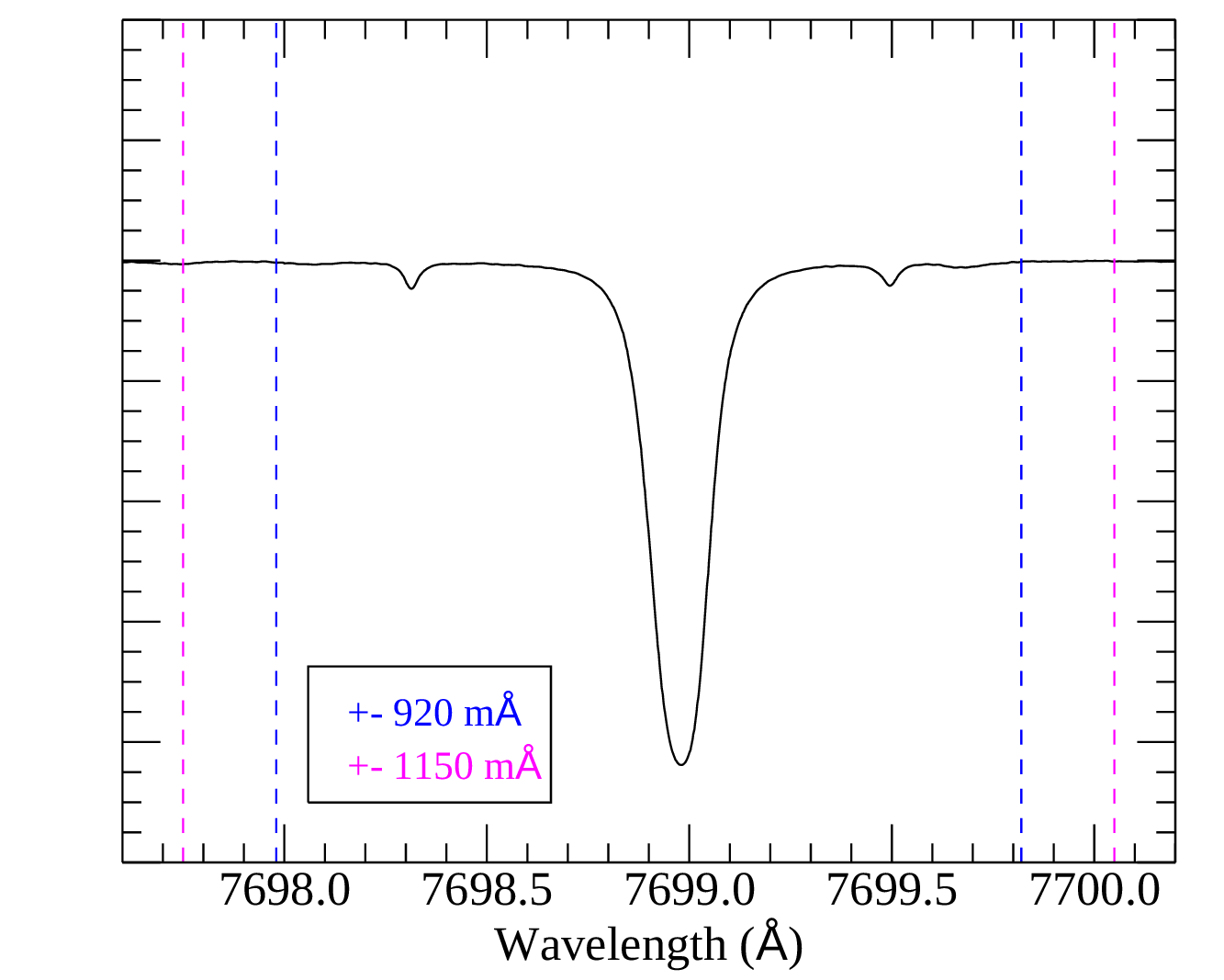}\includegraphics[width=0.33\textwidth]{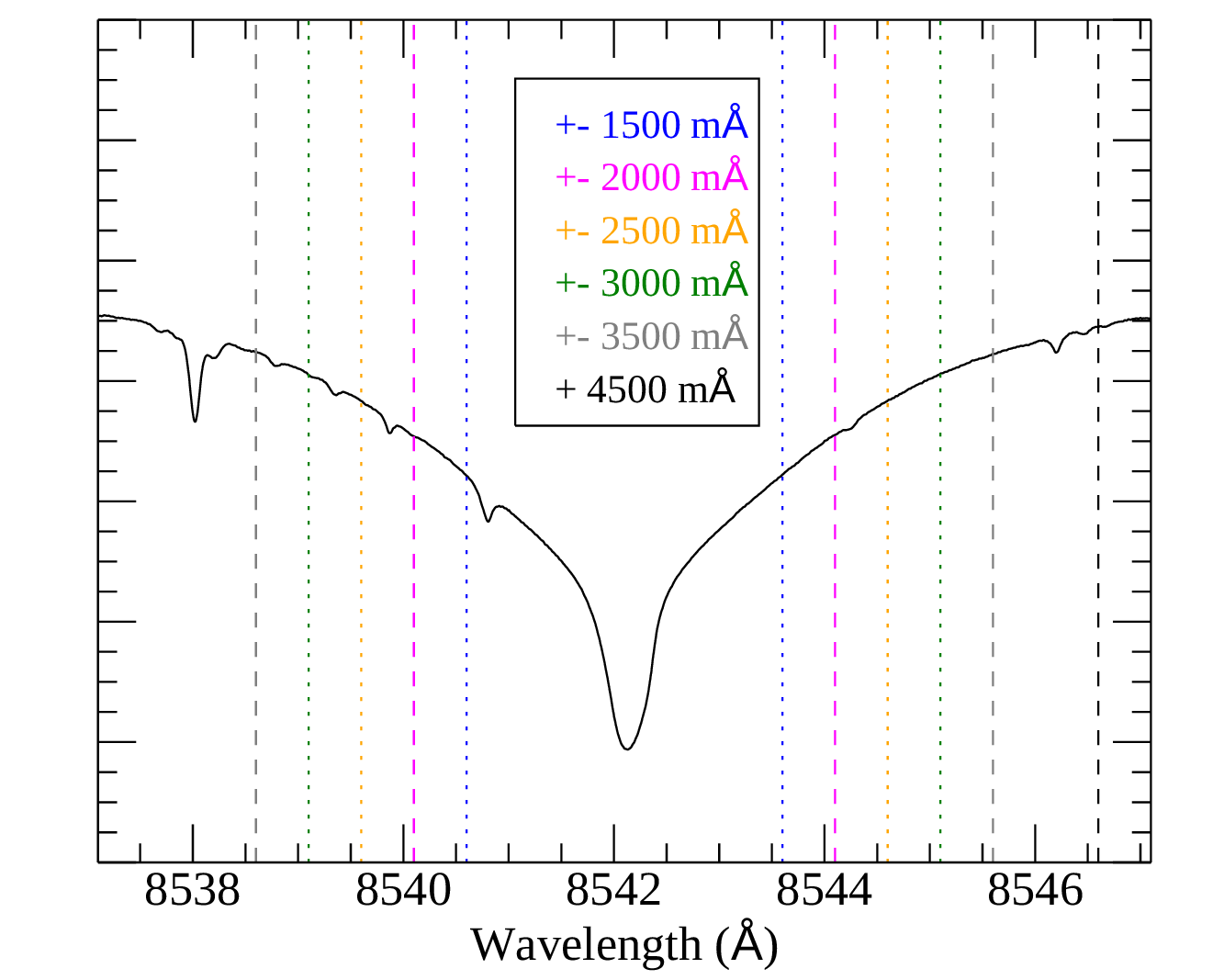}}
    \caption{Chosen wavelength points around the \ion{Fe}{I}~6173~\AA~(left), the \ion{K}{I}~7699~\AA~(middle), and the \ion{Ca}{II}~8542~\AA~line (right) shown on Fourier Transform Spectrometer (FTS) atlas spectral profiles \citep{1999SoPh..184..421N}. The wavelength difference to the line center is indicated in the legend included in each panel. Dashed lines represent wavelength points that are used for both flares presented in this work. Dotted lines represent wavelength points only used for flare~1.}
    \label{fig:cont_points}
\end{figure*}
\indent Following these two categories, WL flares are widely categorized into two classes based on their spectral features and the relationship between intensity increases of the WL and those of hard X-rays (HXRs) or microwaves \citep{1986A&A...159...33M}. Type~I~WL flare spectra exhibit a Balmer jump, extremely strong Balmer line intensities, and are generally coincident (in space and time) with HXR and microwave emission \citep{2016ApJ...816...88K}. Type~II~WL flares, on the other hand, do not cause significant chromospheric Balmer line and/or continuum enhancements and may appear before or after HXR and microwave increases \citep{1995A&AS..110...99F,1999ApJ...512..454D}. The mechanism for type~II~WL flares is assumed to be increased H$^-$~continuum emission from the photosphere \citep{2016ApJ...816...88K}.\\
\indent To distinguish between these mechanisms, we aim to observe the continuum contrast shortward and longward of the Paschen jump (which is situated at $8205.9$~\AA~and stems from the same principles as the Balmer jump). A higher contrast in the blue would strongly indicate recombination radiation \citep{1994ApJ...429..890D}, i.e., a chromospheric origin, whereas a smooth spectrum suggests increased photospheric emission from H$^-$ \citep{1986A&A...159...33M}.


\section{Data and observations}


\subsection{Spectral line selection}

At the Swedish 1-m Solar Telescope \citep[SST;][]{2003SPIE.4853..341S}, we are limited to the available prefilters of the CRisp Imaging SpectroPolarimeter \citep[CRISP;][]{2008ApJ...689L..69S}. For our purposes, this resulted in the selection of the spectral regions that include the \ion{K}{I}~7699~\AA~line (shortward of the jump) and the \ion{Ca}{II}~8542~\AA~line (longward of the jump). We also included the \ion{Fe}{I}~6173~\AA~line as a deep photospheric reference. In the following we give a brief description of each spectral line.\\
\indent The \ion{K}{I}~7699~\AA~line is a resonance line formed in the upper photosphere, just below the temperature minimum region (TMR), at roughly 500~km \citep[e.g.,][]{1975SoPh...43...15D,1992A&A...265..237B,2007AN....328..211H}. This height decreases drastically as the magnetic field concentration increases. It is sensitive to the line-of-sight (LOS) velocity and magnetic field strength \citep[Landé factor $g_\mathrm{eff} \approx 1.3$, see][]{2025ApJ...992..201V} in the upper photosphere and lower chromosphere, which constitute heights where other photospheric lines and the \ion{Ca}{II}~8542~\AA~line do not show similar sensitivities. The intensity is also sensitive to the temperature in lower regions \citep{2017MNRAS.470.1453Q}. Close to the limb ($\mu \leq 0.2$), TMR effects start to become significant \citep{1975SoPh...43...15D}. The continuum around the \ion{K}{I}~line is clean and free of significant other line contributions (see Fig.~\ref{fig:cont_points}, middle panel).\\
\indent The \ion{Ca}{II}~line originates in the lower to mid chromosphere (the height where $\tau = 1$ lies between 0.9 and 1.4~Mm above the surface in the simulations of \citealt{2006ASPC..354..313U}, \citealt{2006ApJ...640.1142P}, and \citealt{2008A&A...480..515C}) and is sensitive to LOS velocity and --- particularly in the photosphere \citep{2008A&A...480..515C} --- temperature variations throughout the entire range from the photosphere to the chromosphere \citep{2017MNRAS.470.1453Q}. As can be seen in the right panel of Fig.~\ref{fig:cont_points}, there is a break in the steepness of the line profile some 30--40~m\AA~away from the line core. In quiet-Sun conditions, the far line wings (beyond that break) are formed in the photosphere \citep{2007ApJ...670..885P,2008A&A...480..515C}, whereas the core regions inside the break are formed in the chromosphere \citep{2008A&A...480..515C}. Under flaring conditions, the sensitivity of the line to the temperature and LOS velocity shifts to lower-lying layers due to the ionization of Ca to \ion{Ca}{III} in higher regions \citep{2018ApJ...860...10K}.\\
\indent The \ion{Fe}{I}~line is formed in the photosphere. It is highly sensitive to the magnetic field due to its large Landé factor \citep[$g_\mathrm{eff} \approx 2.5$, see][]{1977A&A....59..367S}. In quiet Sun conditions, it is an absorption line, and the line core is formed some 200--300~km above the $\tau_{500} = 1$ layer, while the continuum is formed 10--20~km above the solar surface \citep{2006SoPh..239...69N}. The height of line formation can change under flaring conditions, and there can be chromospheric contributions to the line intensity as well \citep{2021ApJ...915...16M,2017ApJ...847...48H,2018ApJ...857L...2H}, which can modify the overall line profile. Similarly, the line can go into emission during flares, indicating strong heating of the lower chromosphere and photosphere. This is more likely to occur above sunspot umbrae \citep{2024AGUFMSH13E2974G}. Similar to the \ion{K}{I}~line, the continuum around the \ion{Fe}{I}~line is clean and free of significant nearby lines (see Fig.~\ref{fig:cont_points}, left panel).


\subsection{Flare observations}

We analyze two events: An M1.8 flare in the active region NOAA 13814 on September 11, 2024 (hereafter flare~1) and an M2.3 flare in the active region NOAA 14140 on \mbox{July 12, 2025} (hereafter flare~2). The observations were carried out with the CRISP instrument on the SST. The CRISP cameras have 2560$\times$2560~pixels, with a pixel size of about 5~$\mu$m \citep{2026A&A...705A..55S}. CRISP has a spatial sampling of 0\farcs044 pixel$^{-1}$ and a roughly circular FOV with 87\arcsec~diameter. The SST adaptive optics system \citep{2024A&A...685A..32S} as well as multi-object multi-frame blind deconvolution \citep[MOMFBD;][]{2005SoPh..228..191V} were used in the data preparation. The final datasets for CRISP utilized in the analysis in this work were generated via the SSTRED reduction pipeline \citep{2015A&A...573A..40D,2021A&A...653A..68L}.\\
\indent Flare~1 was analyzed in detail by \cite{2026A&A...705A.174T}. The location of the center of the FOV was \mbox{$[x_0, y_0] = [145\arcsec, 130\arcsec]$}, i.e., very close to disk center (\mbox{$\mu = 0.98$}). The total observing duration was about 88~min (from 12:30:32~UT to 13:58:34~UT), with a cadence of 36~s. CRISP sampled the \ion{Fe}{I}~6173~\AA~line at 14, the \ion{Ca}{II}~8542~\AA~line at 26, and the \ion{K}{I}~7699~\AA~line at seven spectral points. More information about the choice of continuum spectral points is given in Sec.~\ref{sec:cont_points}. The chosen points are marked with vertical lines in Fig.~\ref{fig:cont_points}.\\
\indent For flare~2, the FOV was centered on \mbox{$[x_0, y_0] = [-852\arcsec, -263\arcsec]$}, meaning close to the eastern limb (\mbox{$\mu = 0.34$}). The total observing duration was about 49~min (from 08:24:48 UT to 09:13:50 UT), with a cadence of 25.5~s. We slightly adjusted our observing program in order to achieve a higher cadence and to increase the spectral sampling of the \ion{K}{I}~line. Therefore, CRISP sampled the \ion{Fe}{I}~line at 14, the \ion{Ca}{II}~line at 20, and the \ion{K}{I}~line at nine spectral points. Spectral points in the \ion{Ca}{II}~line that were only observed during flare~1 are indicated in Fig.~\ref{fig:cont_points} with dotted lines.


\section{Methods}


\subsection{Choice of continuum spectral points}\label{sec:cont_points}
\begin{figure}
    \centering
    \includegraphics[width=0.49\textwidth]{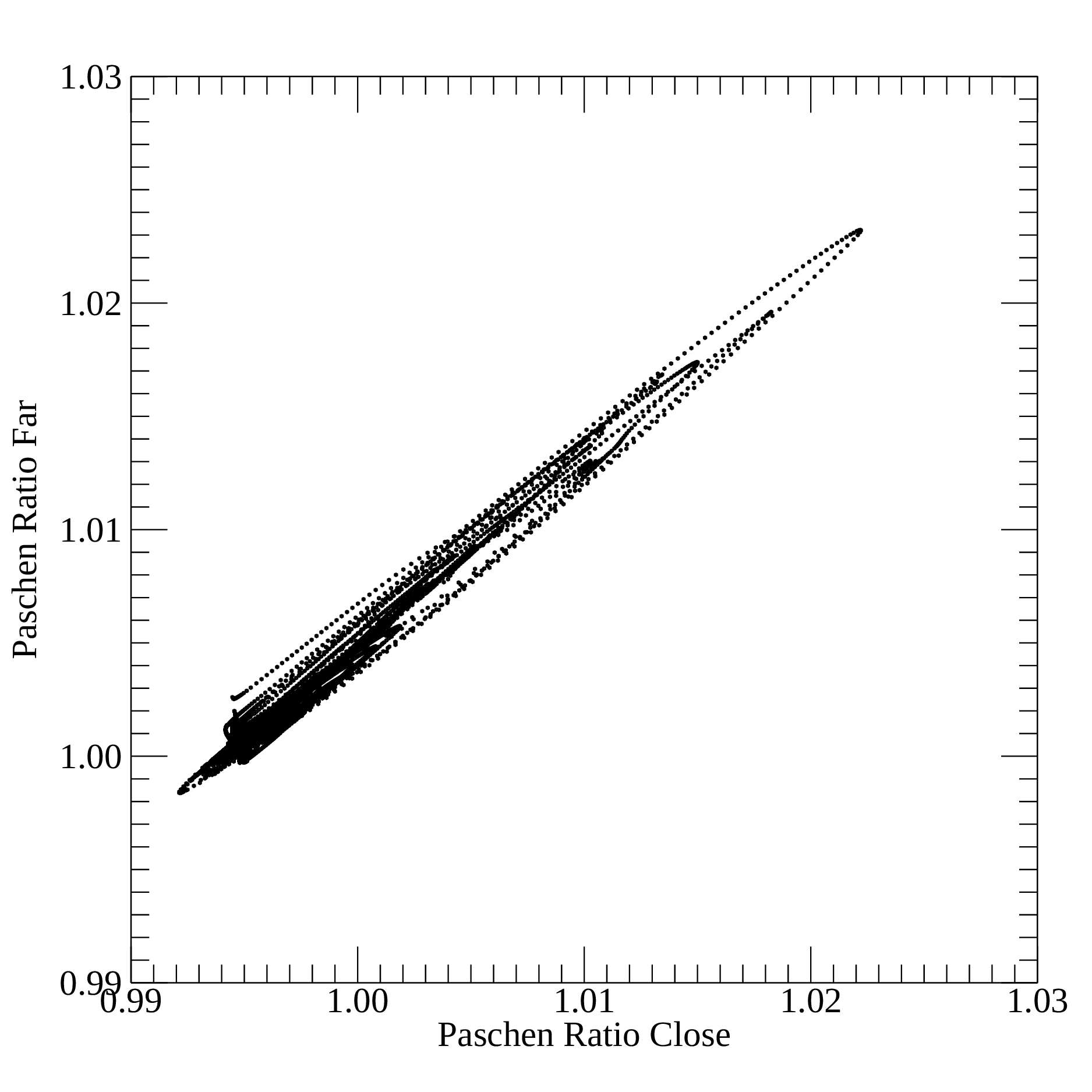}
    \caption{Intensity ratio far from the Paschen jump vs. the intensity ratio close to the Paschen jump for all flare simulations and all times in the F-CHROMA grid. The wavelengths used for the calculation of the faraway ratio are 7800~\AA~and 8760~\AA, whereas those for the close-in Paschen ratio are 8205.7~\AA~and 8206.0~\AA.}
    \label{fig:paschen_ratio_close_far}
\end{figure}
As our main focus in this work is on the continuum emissions, Fig.~\ref{fig:cont_points} contains all (pseudo-) continuum points included in our observations. At the SST, continuum observations around the \ion{Ca}{II}~line are limited by (i) the transmission profile of the prefilter (see Fig.~\ref{fig:prefilter}), (ii) the influence from other lines, and (iii) the stability of CRISP itself, as it can get unstable at certain settings near the extreme ends of the wavelength tuning range (\mbox{P. Sütterlin} 2024, priv. comm.). Taking into account blends with telluric lines (one of which occurs near $-4500$~m\AA~from line center), we are therefore limited to the range \mbox{$[-3500,+4500]$}~m\AA~away from line center. Here, the influence of the line center, taking into account Doppler shifts and line broadening due to temperature increases, is expected to have a very limited effect on the observed intensity. Nonetheless, in this work we call continuum observations that are still influenced by line emission ``pseudo-continuum'' (this is the case for \ion{Ca}{II}). Since the \ion{K}{I}~ and \ion{Fe}{I}~lines are narrow and formed in the photosphere, we do not expect significant shifts due to velocity or temperature changes, and therefore one can expect ``true continuum'' intensities in the spectral points used in this work (see Fig.~\ref{fig:cont_points}).\\
\indent The outermost pseudo-continuum point redward of the \ion{Ca}{II}~line (at 8546.6~\AA) and the continuum point redward of the \ion{K}{I}~line (at 7700.05~\AA) are considerably far away from the wavelength of the Paschen jump (with the bound-free edge at 8205.9~\AA). To estimate whether observations at these wavelengths can be used to infer the Paschen ratio, we made use of the F-CHROMA grid of flare simulations created using the RADYN code \citep{2023A&A...673A.150C}. We define the Paschen ratio as the intensity blueward of the Paschen jump divided by the intensity redward of the jump. In Fig.~\ref{fig:paschen_ratio_close_far}, we show the Paschen ratio calculated taking the closest available points to the jump (at 8205.7~\AA~and 8206.0~\AA) and taking points close to our intended observations (in the F-CHROMA grid, the wavelength points closest to our observations are 7800~\AA~and 8760~\AA, respectively). The figure depicts the ratio for every timestep in every simulation included in the grid. We note here that for our observations, we use the pre-flare-subtracted intensities (i.e., the contrast) rather than the absolute intensities (as are shown in Fig.~\ref{fig:paschen_ratio_close_far}). This is because pre-flare-subtracted intensities in simulations are complicated by intensity decreases that are not apparent in observations. Nonetheless, the evident linear relationship in the figure emboldens the notion that the pseudo-continuum around the \ion{Ca}{II}~line and the continuum around the \ion{K}{I}~line can be used to infer the existence of a Paschen jump. The existence of a jump would be a strong indicator for recombination radiation from the chromosphere as the dominant mechanism \citep{1994ApJ...429..890D}.


\subsection{(Hard) X-ray data}

To investigate the flare heating mechanisms, we analyze the relationship between thermal and nonthermal X-ray emissions, i.e., between soft X-ray (SXR) and HXR enhancements. The time derivative of the SXR corresponds well with HXR increases in situations of chromospheric evaporation due to electron-beam precipitation \citep{1999ApJ...514..472M}. This is called the Neupert effect \citep{1968ApJ...153L..59N,2002A&A...392..699V}.\\
\indent In this work, we make use of full-disk SXR data taken by the Geostationary Operational Environmental Satellites (GOES). GOES provides measurements in two broadband channels, \mbox{0.5--4~\AA}~and 1--8~\AA, using ion chamber detectors \citep{1994SoPh..154..275G}. The cadence is 1~s. The 1--8~\AA~channel is the one that fits our purpose best, as it is the channel generally used for studies of the Neupert effect. We supplement the GOES measurements with full-disk HXR spectral data taken by the Hard X-ray Imager \citep[HXI;][]{2019RAA....19..160Z} onboard the Advanced Space-based Solar Observatory \citep[ASO-S;][]{2019RAA....19..156G} in order to compare the proxy HXR with true HXR measurements. HXI provides spectra and images in the energy range of 10--300 keV with an energy resolution of about 16.6\% at 32 keV and a temporal resolution of 4~s in regular mode down to 0.125~s in burst mode \citep{2024SoPh..299..153S}. HXI has been used previously by \cite{2024SoPh..299...57L} to investigate the Neupert effect in solar flares.\\
\indent For the event to be consistent with the Neupert effect, the end of HXR emission needs to lie within 1 min of the maximum of SXR emission \citep[following the definition of][]{2002A&A...392..699V}. If the maximum of SXR enhancements occurs later, this can indicate another heating mechanism at play other than electron-beam precipitation.
\begin{figure*}
    \resizebox{\hsize}{!}
    {\includegraphics[width=0.99\textwidth]{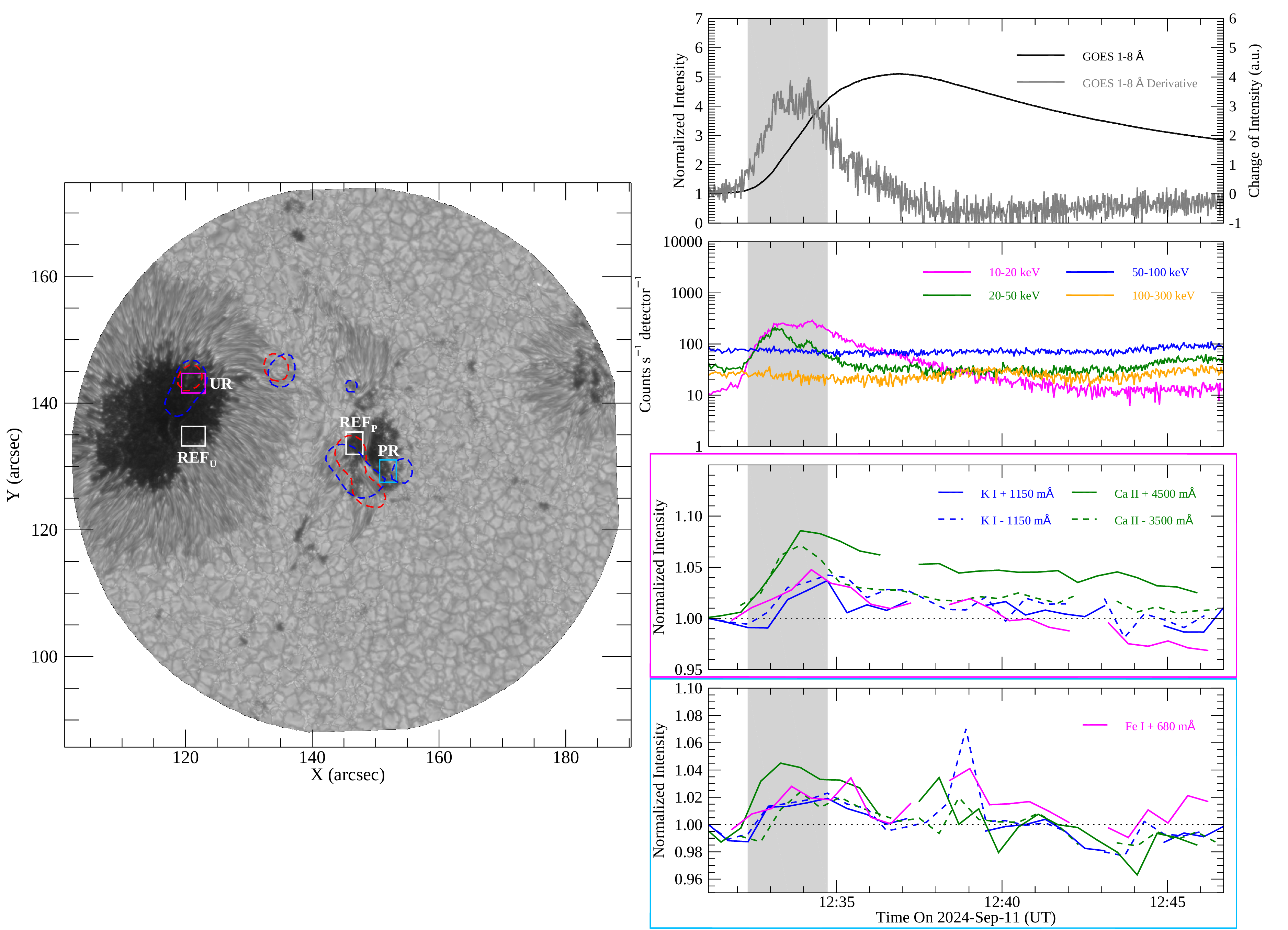}}
    \caption{Relative timing of white-light enhancements in flare~1 observed by the SST and X-ray increases observed by GOES and ASO-S HXI. Left: SST observation of the continuum around the \ion{K}{I}~7699~\AA~line (1150~m\AA~longward of line center) taken at 12:32:55~UT. The cyan and magenta rectangles correspond to the fields of view taken for the light curves shown on the right. The white rectangles indicate the regions used as reference for general intensity changes. All regions are additionally marked with letters (UR for umbral region, PR for pore region, as well as REF$_\mathrm{U}$ and REF$_\mathrm{P}$ for reference regions in the umbra and pore, respectively). The regions experiencing the highest HXR increases (60-\% contours) between 12:32:28~UT and 12:34:30~UT are indicated with blue (15-20~keV) and red (20-40~keV) dashed contours. First panel on the right: Normalized intensity as a function of time for the \mbox{GOES 1--8~\AA~(black)} and its derivative (gray). Second panel on the right: Hard X-ray measurements taken by HXI onboard ASO-S between 10 and 20 keV (magenta), 20 and 50 keV (green), 50 and 100 keV (blue), and 100 and 300 keV (orange). Bottom two panels on the right: Normalized intensity of the (pseudo-) continuum around \ion{K}{I} (blue), \ion{Ca}{II} (green), and \ion{Fe}{I} (magenta) taken by the SST averaged over the two subregions UR and PR. Solid (dashed) lines mark the red (blue) (pseudo-) continuum. Measurements with bad seeing have been removed (see Appendix). The horizontal dotted line in these two panels marks the intensity at the start of our observations. The shaded gray area in each panel shows the time of observed WL enhancements.}
    \label{fig:goes_sst_1}
\end{figure*}


\subsection{Detection of white-light enhancements}\label{sec:wl_detection}

Due to the fact that the WL enhancements in the observed flares are very low, it is not possible to identify excess emission via a simple threshold technique. This is underlined by the fact that the intensity changes and motions induced by both the granulation and the movement of bright features may result in very large differential pixel values, without the influence of the flare itself. We therefore manually identified WL pixel candidates by visual inspection of the series of images and subsequently confirmed our initial estimates via examination of the temporal evolution of the light curves (in the \ion{K}{I} continuum) of each WL pixel candidate. We only retained those pixels that displayed distinct, impulsive intensity increases unrelated to seeing effects.


\section{Results}

\begin{figure*}
    \resizebox{\hsize}{!}
    {\hspace{1.6cm}\includegraphics[width=1.0\hsize]{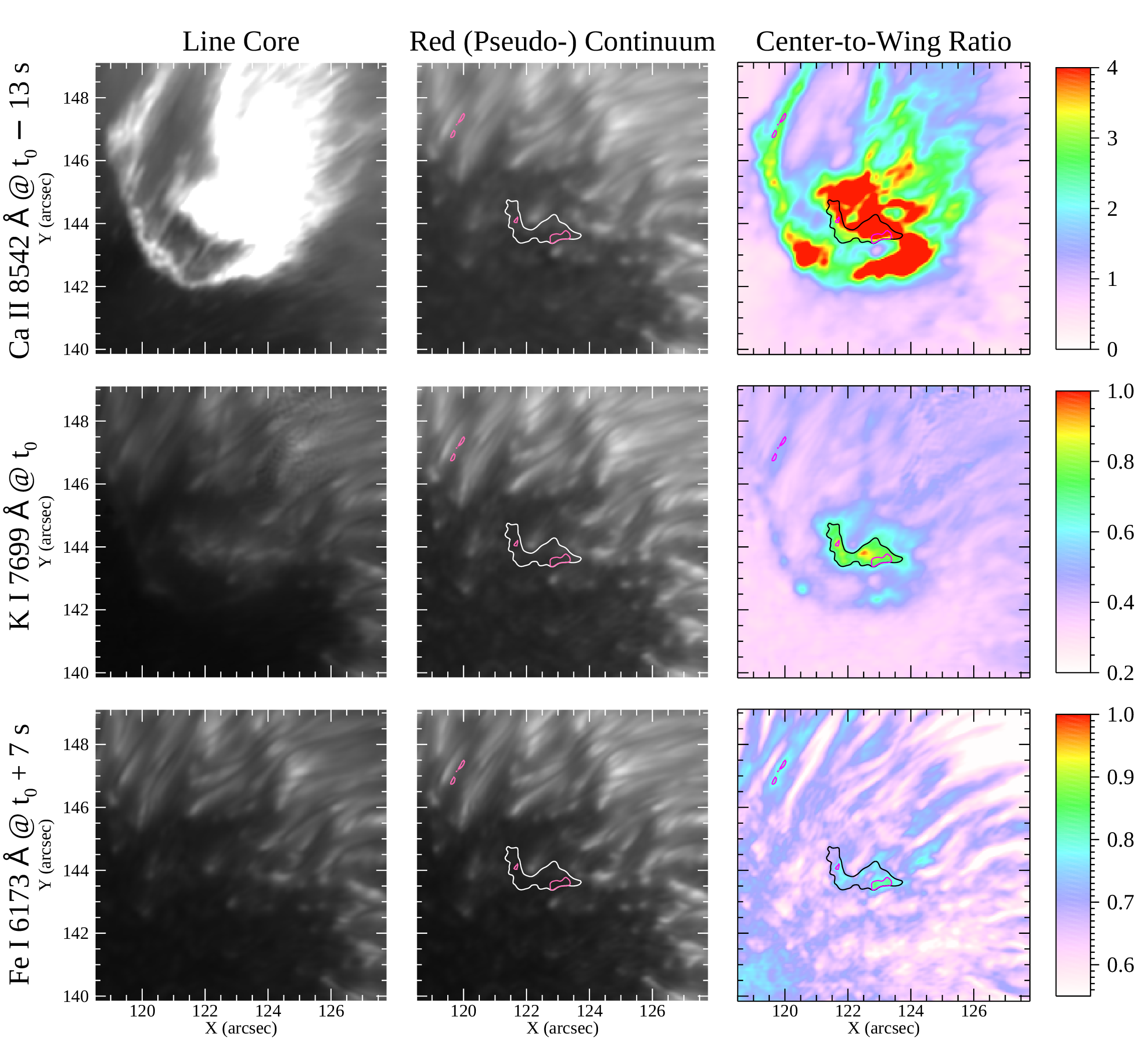}\hspace{1.6cm}}
    \caption{SST observations of the \ion{Ca}{II}~8542~\AA~(top row), the \ion{K}{I}~7699~\AA~(middle row), and the \ion{Fe}{I}~6173~\AA~line (bottom row) for flare~1. The left column shows the line core, the middle column the red-most (pseudo-) continuum point, and the right column the center-to-wing ratio (CWR). Colorbars for the CWR for each row are given on the right edge. The white (black) contours in the middle (right) column mark the regions where the CWR exceeds 0.6 for the \ion{K}{I}~7699~\AA~line, whereas magenta contours demarcate where the CWR for the \ion{Fe}{I}~6173~\AA~line exceeds 0.8. Note the time difference between the measurements of each line as indicated on the left. The reference time $t_0$ is 12:33:31~UT.}
    \label{fig:flare_general_cwr_1}
\end{figure*}


\subsection{M1.8 flare on September 11, 2024}


\subsubsection{General description}\label{sec:flare_1}

The M1.8~flare as measured by the GOES satellites started at 12:27, peaked at 12:36, and ended at 12:47 (all times in UT). Our SST observations cover the impulsive and gradual phases but miss the initial onset. The flare appeared close to disk center ($\mu = 0.98$). As can be seen in the upper two panels of Fig.~\ref{fig:goes_sst_1}, the event is consistent with the Neupert effect since the derivative of SXR enhancements coincides temporally with HXR emission. The left half of Fig.~\ref{fig:goes_sst_1} contains a \ion{K}{I}~continuum image of the whole FOV of CRISP. Morphologically, the flare happened between a fully developed sunspot, a proto-sunspot (for simplicity we call this a pore from now on, although that designation is not strictly correct) with filamentary structures surrounding its umbra and a lightbridge in its center (separating it into an eastern and a western part), as well as the decaying remnants of another sunspot (to the west). The lightbridge in the pore marks the neutral line between two opposing magnetic polarities as seen in the magnetogram derived from the photospheric \ion{Fe}{i} line (not shown).\\
\indent Our observations show that the chromospheric response is substantial, with the \ion{Ca}{II}~8542~\AA~line core transitioning into full emission across most of the flaring region (Fig.~\ref{fig:flare_general_cwr_1}). To visualize the photospheric response, we examine the center-to-wing ratio (CWR), i.e., the ratio of line-core intensity to the intensity at the wavelength position farthest away from the line center. This is depicted in the right column of Fig.~\ref{fig:flare_general_cwr_1}. While the \ion{Fe}{I}~6173~\AA~line generally remains in absorption, the \ion{K}{I}~7699~\AA~line exhibits emission cores in localized patches, indicating significant heating penetrating into the upper photosphere.\\
\begin{figure*}
    \resizebox{1.0\hsize}{!}
    {\hspace{0.5cm}\includegraphics[width=1.0\hsize]{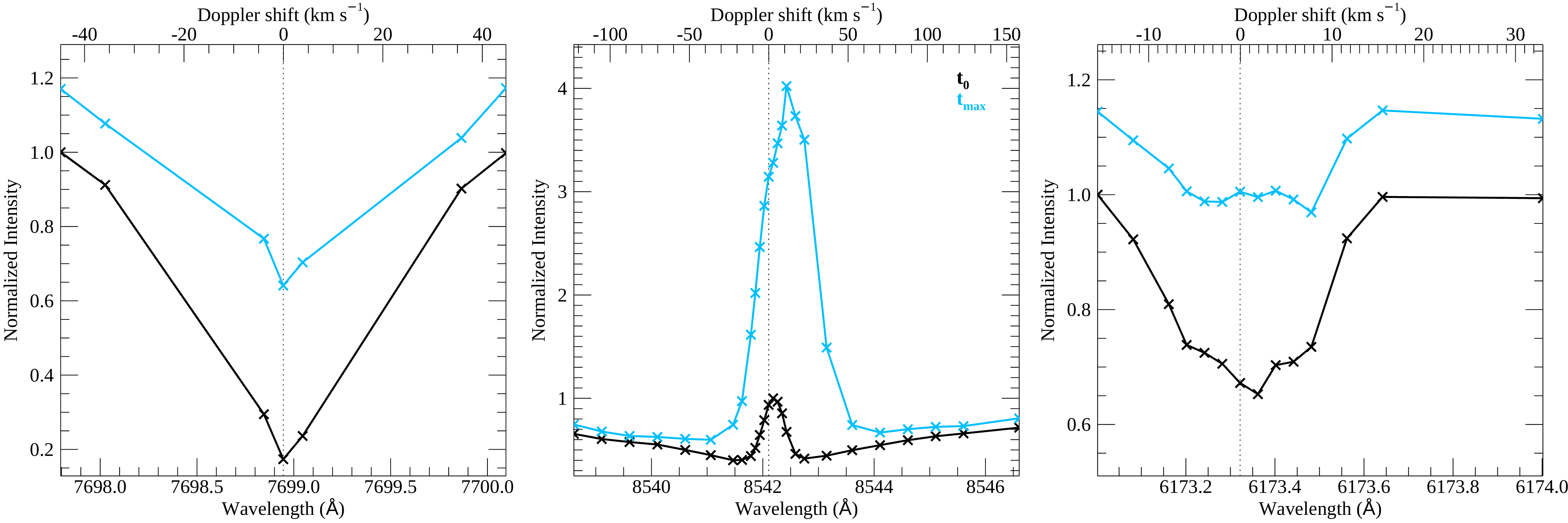}\hspace{0.5cm}}
    \caption{Example spectral profiles for flare~1. Left panel: \ion{K}{I}~7699~\AA~profiles. Middle panel: \ion{Ca}{II}~8542~\AA~profiles. Right panel: \ion{Fe}{I}~6173~\AA~profiles. In all panels, black profiles correspond to the start time of our observations, whereas cyan profiles were taken at the time of maximum WL enhancements. Crosses mark the measured wavelength points. The vertical dotted line in each panel marks the location of the unshifted line core.}
    \label{fig:profiles_1}
\end{figure*}
\indent We classified the spectral profiles at the time of maximum WL enhancement into distinct phenomenological categories (one example per line is displayed in Fig.~\ref{fig:profiles_1}, and further examples are shown in the appendix). \ion{Ca}{II}~8542~\AA~profiles are dominated by strong emission cores. We identify signatures consistent with chromospheric evaporation (blueshifts up to 20~km~s$^{-1}$) and condensation (redshifts up to 50~km~s$^{-1}$), as well as complex dual-component profiles \citep[cf.][]{2020ApJ...895....6G}. The majority of pixels show the \ion{K}{I}~7699~\AA~line in absorption with reduced depth (shallower cores). A subset of pixels, particularly within the pore (see the bottom-left panel in Fig.~\ref{fig:profiles_ki}), exhibits central emission cores similar to the reversal seen in \ion{Ca}{II}, albeit weaker. Asymmetries in the wings often mimic redshifts, suggesting downward velocities in the formation region. \ion{Fe}{I}~6173~\AA~profiles remain in absorption but display complex distortions. Common features include ``W-shaped'' central reversals \citep[as in][]{2018ApJ...857L...2H}, indicating mid-photospheric heating, and strong asymmetries. Notably, we observe profiles where the red wing extends significantly above the pre-flare continuum level, potentially indicating a moving heating front or strong velocity gradients. In summary, redshifts and red-wing asymmetries are pervasive across all three lines, suggesting that the WL-emitting regions are dominated by downward-moving, dense material (condensation) compressing the lower atmosphere.


\subsubsection{White-light enhancements}

\begin{figure*}
    \resizebox{\hsize}{!}
    {\includegraphics[width=0.99\textwidth]{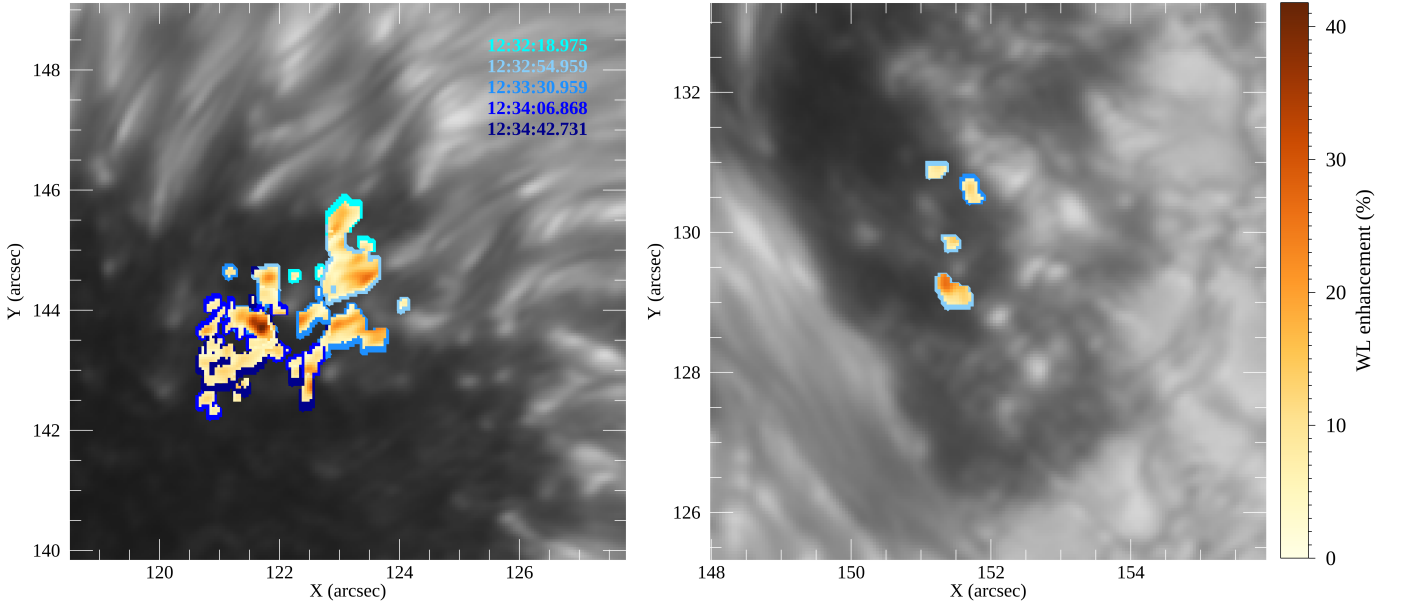}}
    \caption{\ion{K}{I} continuum intensity enhancement (in percent) for each identified WL pixel for flare~1 overlaid on images of the \ion{K}{I} continuum. The left panel shows a subregion of the sunspot, while the right panel depicts a zoom into the western part of the pore. Colored contours around the WL areas mark the time of maximum WL enhancement, as indicated in the top-right corner of the left panel.}
    \label{fig:wl_pixels_1}
\end{figure*}
\begin{figure}
    \centering
    \includegraphics[width=0.49\textwidth]{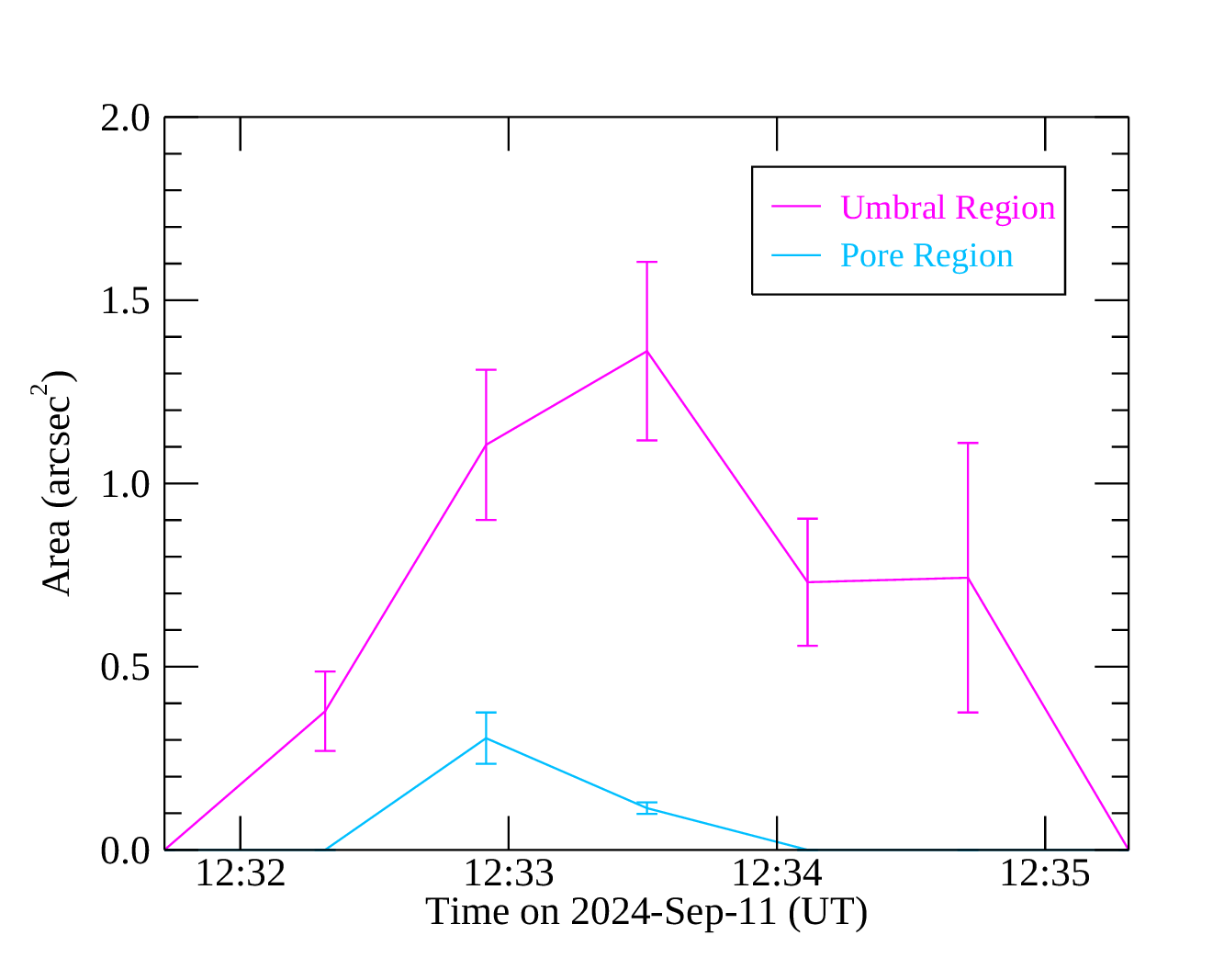}
    \caption{Temporal evolution of the WL area (in the \ion{K}{I} continuum) for flare~1. WL enhancements in the umbral region are shown with magenta color, those in the pore region with cyan.}
    \label{fig:wl_area_1}
\end{figure}
\begin{figure}
    \centering
    \includegraphics[width=0.49\textwidth]{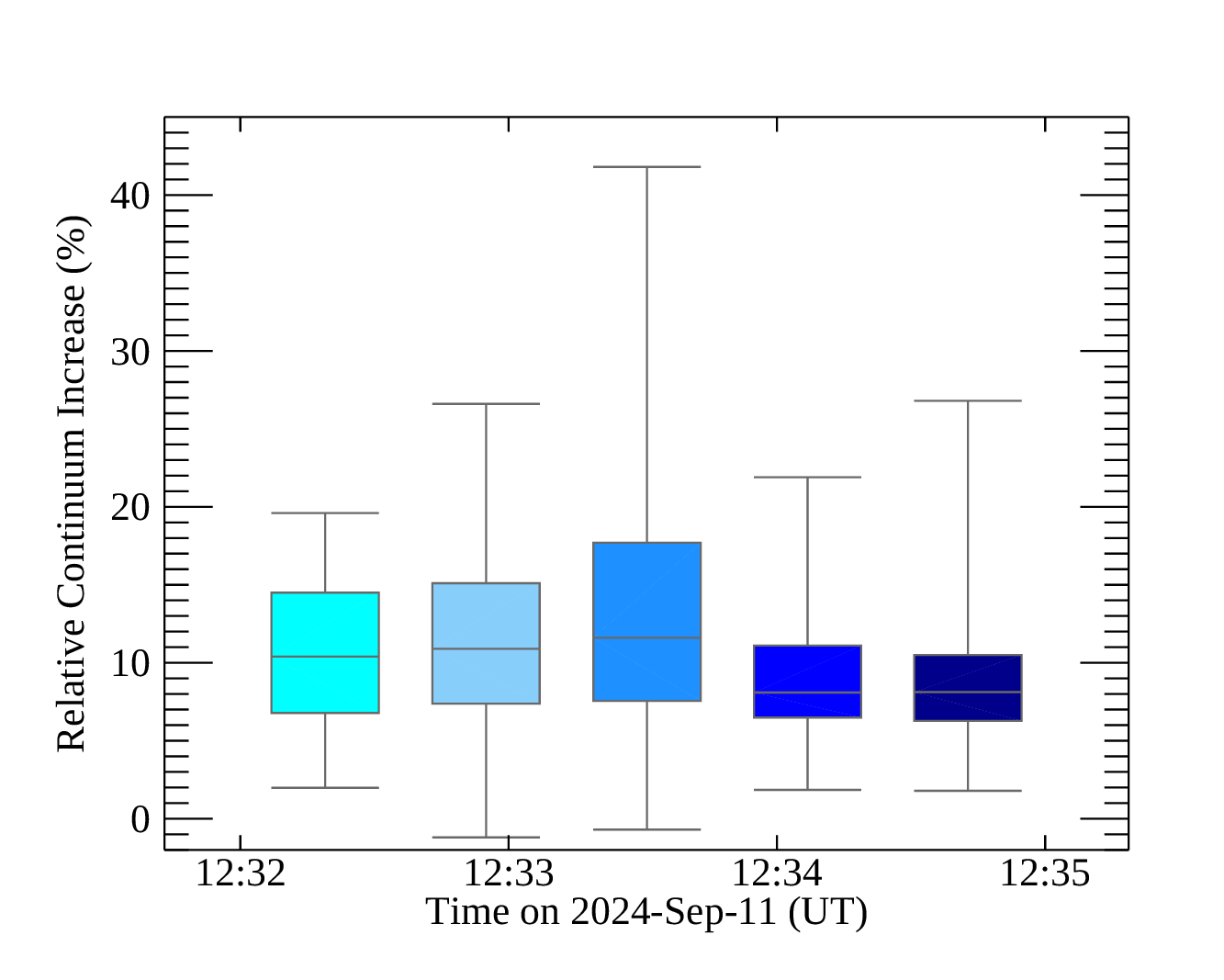}
    \caption{Statistical characteristics of the WL excess intensity (in the \ion{K}{I} continuum) in flare~1 as a function of time. The color of each box corresponds to the colors in Fig.~\ref{fig:wl_pixels_1}. For each timestep, the horizontal gray line indicates the median excess, the box represents the interquartile range (25th--75th percentile), and the whiskers extend to the minimum and maximum values.}
    \label{fig:wl_intensity_box_1}
\end{figure}
\begin{figure}
    \centering
    \includegraphics[width=0.49\textwidth]{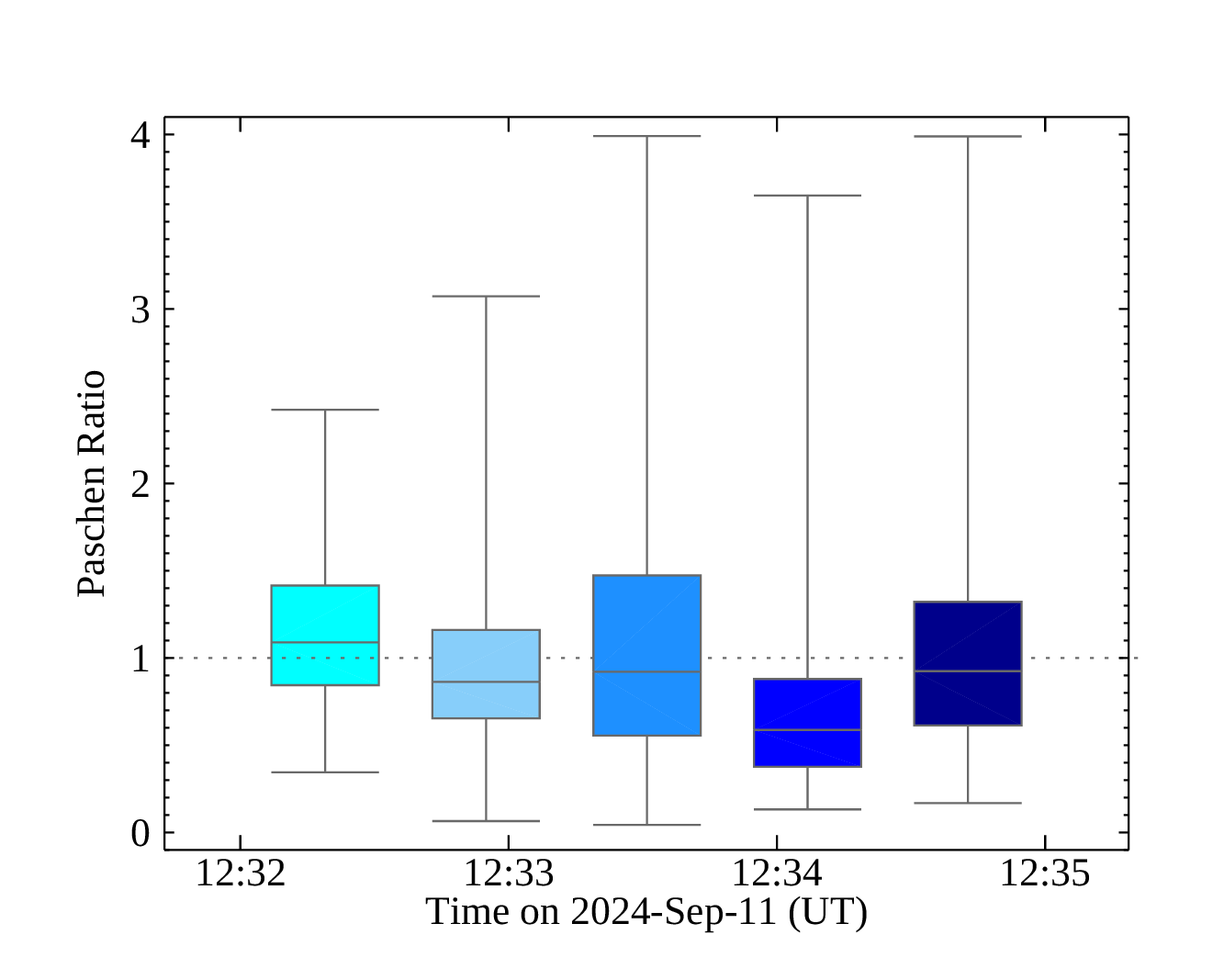}
    \caption{Distribution of the Paschen ratio (defined as the ratio of the excess intensity at 7700~\AA~and at 8546.6~\AA) across WL pixels (following the same style as Fig.~\ref{fig:wl_intensity_box_1}). The color of each box corresponds to the colors in Fig.~\ref{fig:wl_pixels_1}. Paschen ratio unity is marked with a dotted horizontal line.}
    \label{fig:paschen_ratio_1}
\end{figure}

Continuum enhancements are clearly detected in the \ion{Ca}{II} pseudo-continuum and, more faintly, in the \ion{K}{I} and \ion{Fe}{I} true continua. These WL kernels are spatially co-located with the brightest chromospheric footpoints in the northwestern boundary of the penumbra and the umbra (we call this the umbral region, UR) and the western part of the pore (pore region, PR).\\
\indent The bottom two panels of the right half of Fig.~\ref{fig:goes_sst_1} show subregion light curves of the areas indicated with colored boxes in the left half of the figure. The values are normalized to the pre-flare level. Finally, we applied detrending taking into account (i) the regions within the white boxes in the figure (which are areas similar to the UR and PR, but without emission enhancements), and (ii) the (approximately linear) increase due to bright features moving into the respective area. The subregion light curve for the UR in the continuum points around our selected lines shows short-lived increases lasting about 2.5~min and displaying peak increases (with respect to the start of our observations) of 3.7\% (\ion{K}{I}~+1150~m\AA), 8.6\% (\ion{Ca}{II}~+4500 m\AA), and 4.8\% (\ion{Fe}{I}~+680~m\AA). These increases are co-temporal with those seen in the time derivative of the GOES SXR as well as with HXR increases in the 10--20 keV and 20--50 keV ranges. In the PR, the increases are less pronounced. The peak WL enhancements in this region are 1.6\% (\ion{K}{I}~+1150~m\AA), \mbox{7.3\% (\ion{Ca}{II}~+4500 m\AA)}, and 0.6\% (\ion{Fe}{I}~+680~m\AA). We note here that the gray-shaded areas in Fig.~\ref{fig:goes_sst_1}, which signify the time of WL enhancements, are based on our investigation of the lightcurves of each pixel (see Sec.~\ref{sec:wl_detection}), and not on the subregion light curves.\\
\indent The spatial correlation between HXR (inferred using the CLEAN algorithm) and WL enhancements is strong in the UR, but weak in the PR, where an offset of several arcseconds is apparent. 60-\% contours of the HXR intensity are overlaid onto the SST image in Fig.~\ref{fig:goes_sst_1}. We note that these images are based on a beta version of the HXI software used for these purposes, and further improvements are currently ongoing (Y. Su 2026, priv. comm.).\\
\indent The empirical identification of WL pixels leads to the WL pixel map shown in Fig.~\ref{fig:wl_pixels_1}. Both the intensity increases and the background images represent the \ion{K}{I} continuum. The map is color-coded according to the time at which a given pixel showed its maximum WL enhancement. A clear front of WL emission traveling toward the center of the sunspot umbra is visible. From the overlaid WL intensity image it can be seen that WL emission is not uniform in space, as clear kernels can be observed. WL enhancements in the PR (right panel of Fig.~\ref{fig:wl_pixels_1}) are generally smaller than in the UR. Corresponding images for the \ion{Ca}{II} pseudo-continuum and the \ion{Fe}{I} continuum are included in the appendix. They show similar kernels of WL emission, albeit slightly different in location to those in the \ion{K}{I} continuum, possibly owing to the uncertainty in intensity due to the applied interpolation (to the time of the respective observation in the \ion{K}{I} continuum).\\
\indent The temporal evolution of the WL area (per definition in the \ion{K}{I} continuum) in each region is depicted in Fig.~\ref{fig:wl_area_1}. Using CRISP's 0\farcs044 pixel$^{-1}$ resolution, together with the number of pixels identified as WL pixels, we arrive at a peak WL area of $1.36 \pm 0.24$~arcsec$^2$ in the UR and $0.30 \pm 0.07$~arcsec$^2$ in the PR, respectively. The total WL area (i.e., the union of all WL masks over time) in the UR (PR) is $3.58 \pm 0.74$~arcsec$^2$ ($0.42 \pm 0.09$~arcsec$^2$).\\
\indent As Fig.~\ref{fig:wl_intensity_box_1} shows, the median WL excess (in the \ion{K}{I} continuum) is roughly 10\% for each timestep. The excess is slightly higher for the \ion{Ca}{II} pseudo-continuum and generally much lower for the \ion{Fe}{I} continuum (see appendix). The excess in the \ion{K}{I} continuum has a maximum at 12:33:30~UT. This could be due to either (i) a strengthening of the operating mechanism(s), or (ii) a higher contrast, since the WL enhancements are seen closer to the center of the sunspot as time progresses, where the umbral background intensity is lower.\\
\indent In Fig.~\ref{fig:paschen_ratio_1}, we show the Paschen ratio for every WL pixel. For this, we interpolated the \ion{Ca}{II}~pseudo-continuum measurements to the time of the corresponding \ion{K}{I}~continuum observation. The median of the distribution (signified by the horizontal black line within each colored rectangle) for each timestep (with the exception of the first one) --- and therefore the Paschen ratio for most WL pixels --- is below unity.


\subsection{M2.3 flare on July 12, 2025}


\subsubsection{General description}\label{sec:flare_2}

\begin{figure*}
    \resizebox{\hsize}{!}
    {\includegraphics[width=0.99\textwidth]{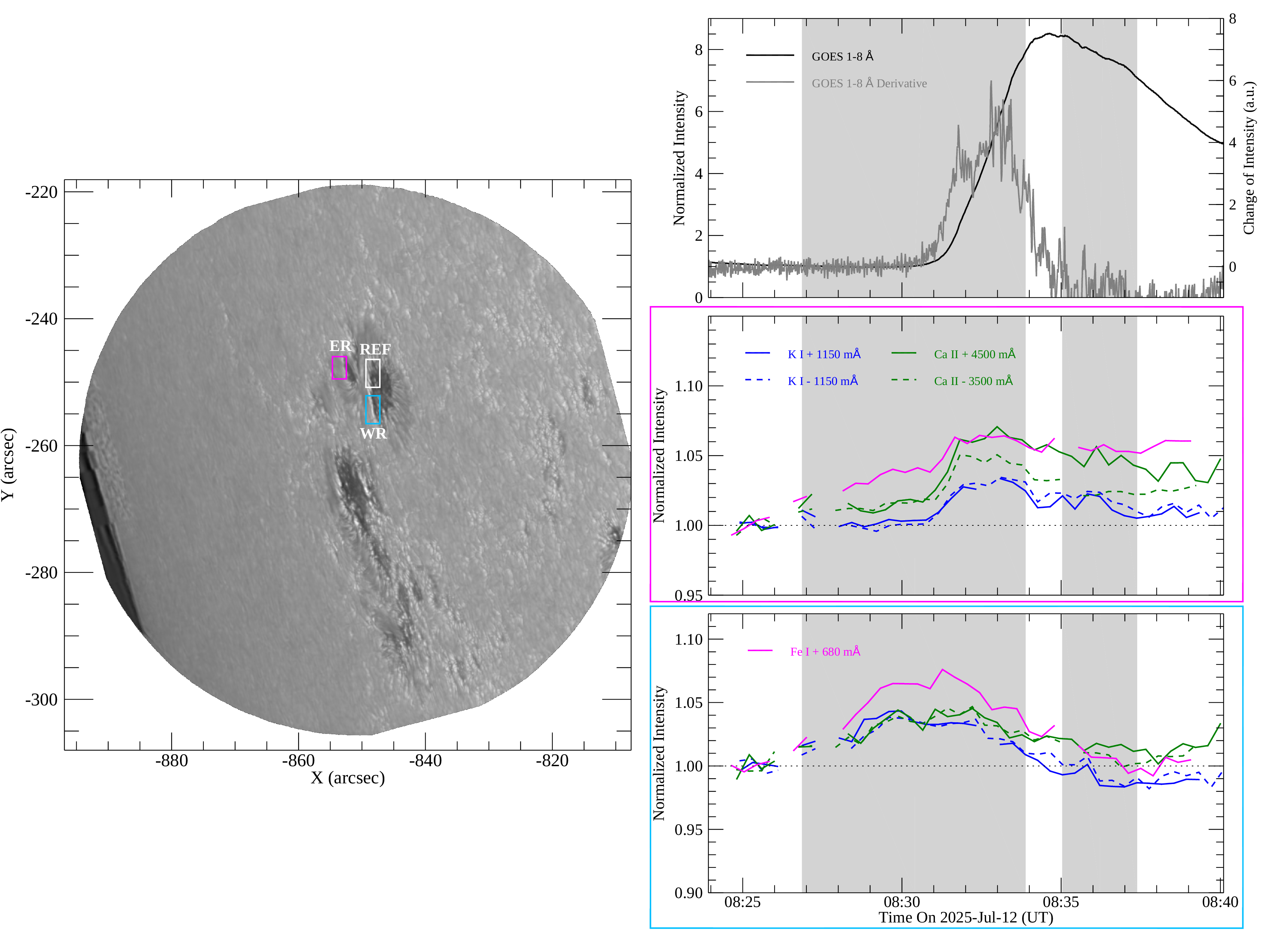}}
    \caption{Relative timing of WL enhancements in flare~2 (following the same style as Fig.~\ref{fig:goes_sst_1}). The SST image was taken at 08:26:29~UT. The region designations are ER (eastern region), WR (western region), and REF (reference region). No reliable ASO-S HXI measurements were available for this flare.}
    \label{fig:goes_sst_2}
\end{figure*}
\begin{figure*}
    \resizebox{\hsize}{!}
    {\hspace{1.6cm}\includegraphics[width=1.0\hsize]{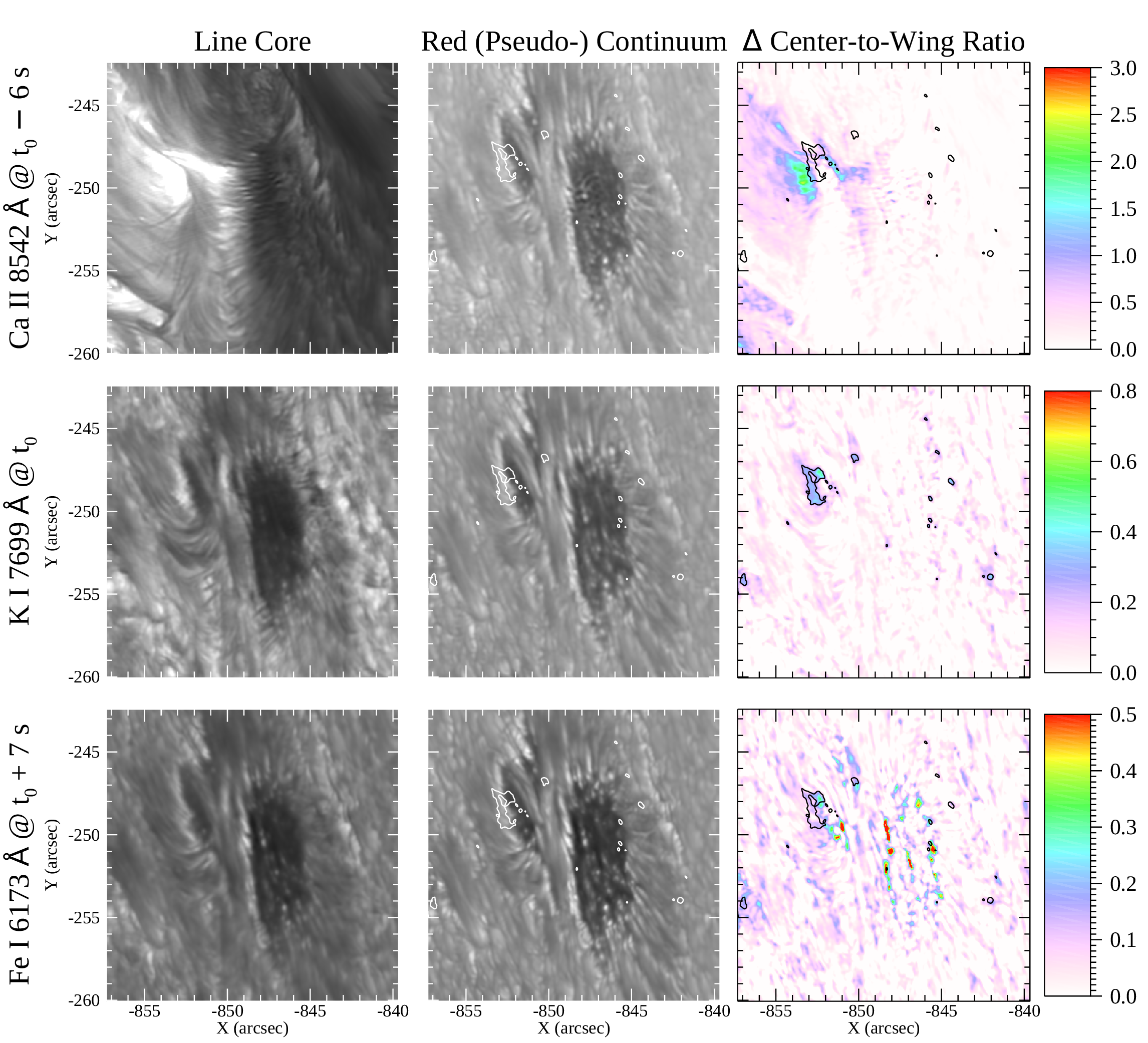}\hspace{1.6cm}}
    \caption{SST observations of flare~2 (following the same style as Fig.~\ref{fig:flare_general_cwr_1}). The right column depicts the change of the CWR relative to the start of our observations. No contours for the CWR of \ion{Fe}{I} have been drawn. The reference time $t_0$ is 08:33:52~UT.}
    \label{fig:flare_general_cwr_2}
\end{figure*}
\begin{figure*}
    \resizebox{1.0\hsize}{!}
    {\hspace{0.5cm}\includegraphics[width=1.0\hsize]{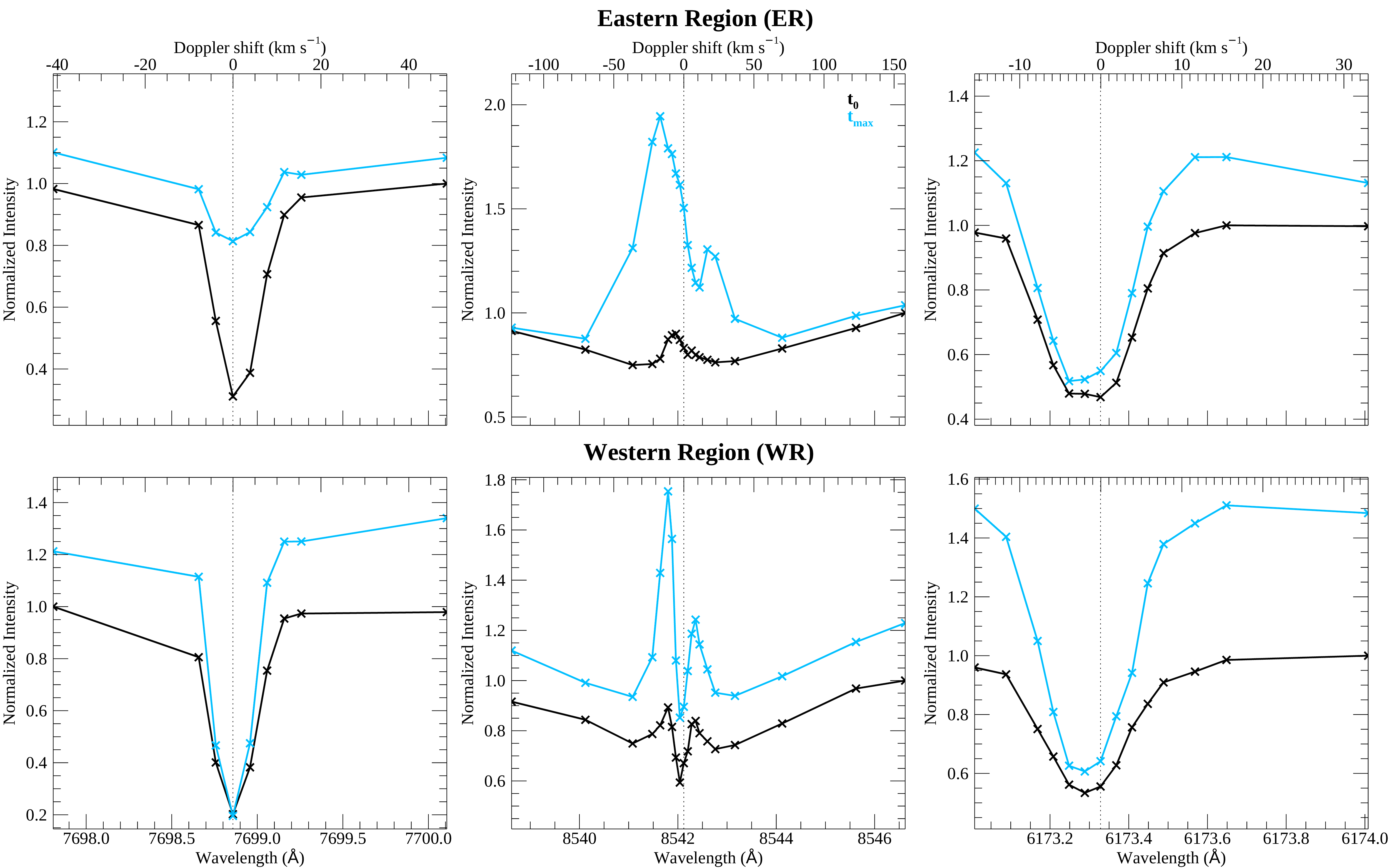}\hspace{0.5cm}}
    \caption{Example spectral profiles for flare~2 (following the same style as Fig.~\ref{fig:profiles_1}). The top row shows example profiles for the eastern region (ER), the bottom row for the western region (WR).}
    \label{fig:profiles_2}
\end{figure*}

The second flare under investigation is an M2.3 flare on \mbox{July 12, 2025}. It followed shortly (6~min) after a smaller C4.8 flare at the same location, which was near the limb (\mbox{$\mu \approx 0.34$}). The flare occurred close to the northeastern sunspot of the sunspot group (see Fig.~\ref{fig:goes_sst_2}, left panel). The X-ray event started at 08:29, peaked at 08:34, and ended at 08:42 (all times in UT). The GOES SXR and its derivative as well as WL subregion light curves are shown in Fig.~\ref{fig:goes_sst_2}. Unfortunately, no reliable ASO-S HXI measurements are available for this time due to the passage through the South Atlantic Anomaly from 08:12~UT to 08:29~UT and the satellite being in the Earth's shadow thereafter until 08:43~UT. We therefore rely on the GOES SXR derivative as the sole proxy for nonthermal emission.\\
\indent The signatures in the \ion{Ca}{II}~line are not as pronounced in this flare as in flare~1. We notice filamentary material in the foreground blocking some of the flare emission in the line core. We observe \ion{K}{I} emission cores spatially coincident with the base of the extensive \ion{Ca}{II} fibril structures. Similar to flare~1, the signatures are stronger in \ion{K}{I} than in \ion{Fe}{I}. Furthermore, the \ion{K}{I} emissions exhibit a clear elongated structure. Combined with the limb viewing angle, this leads us to believe that these emissions extend into higher-lying layers.\\
\indent The spectral response at WL maximum (some examples of which are shown in Fig.~\ref{fig:profiles_2}) shows a clear dichotomy between the eastern region (ER) and western region (WR). In the ER, the \ion{K}{I} and \ion{Fe}{I} profiles are characterized by blueshifts ($\sim$10~km~s$^{-1}$) and red-wing asymmetries, often retaining a central absorption core. \ion{Ca}{II} profiles display double-peaked emission (where the blue peak is dominant) with a central reversal, consistent with optically thick source functions that decrease with height above a local maximum. The WR exhibits peculiar ``deep absorption'' profiles. Both \ion{K}{I} and \ion{Fe}{I} show deep absorption cores (comparable to pre-flare depths) but with a significantly elevated continuum level. In \ion{Ca}{II}, the profiles show complex, W-shaped central emission features within a broad absorption trough. These spectral signatures suggest that while the continuum-forming layers are heated, a substantial column of cool, absorbing plasma remains above the flare site, partially masking the chromospheric emission lines. Overall, the photospheric lines (\ion{Fe}{I} and \ion{K}{I}) are less distorted than in flare~1, likely due to the limb geometry.


\subsubsection{White-light enhancements}

\begin{figure}
    \centering
    \includegraphics[width=0.48\textwidth]{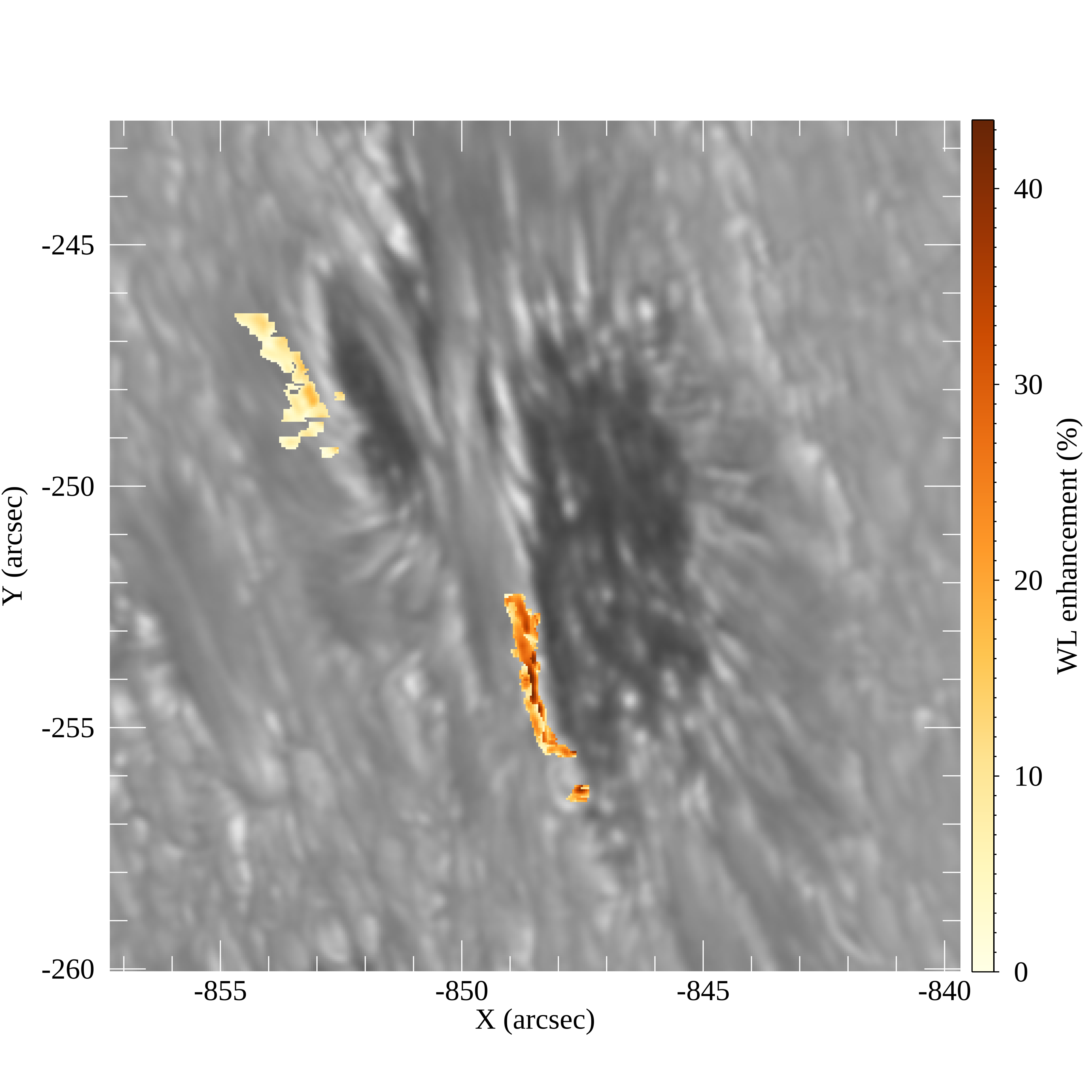}
    \caption{\ion{K}{I} continuum intensity enhancement (in percent) for each identified WL pixel for flare~2.}
    \label{fig:wl_pixels_2}
\end{figure}

\begin{figure}
    \centering
    \includegraphics[width=0.49\textwidth]{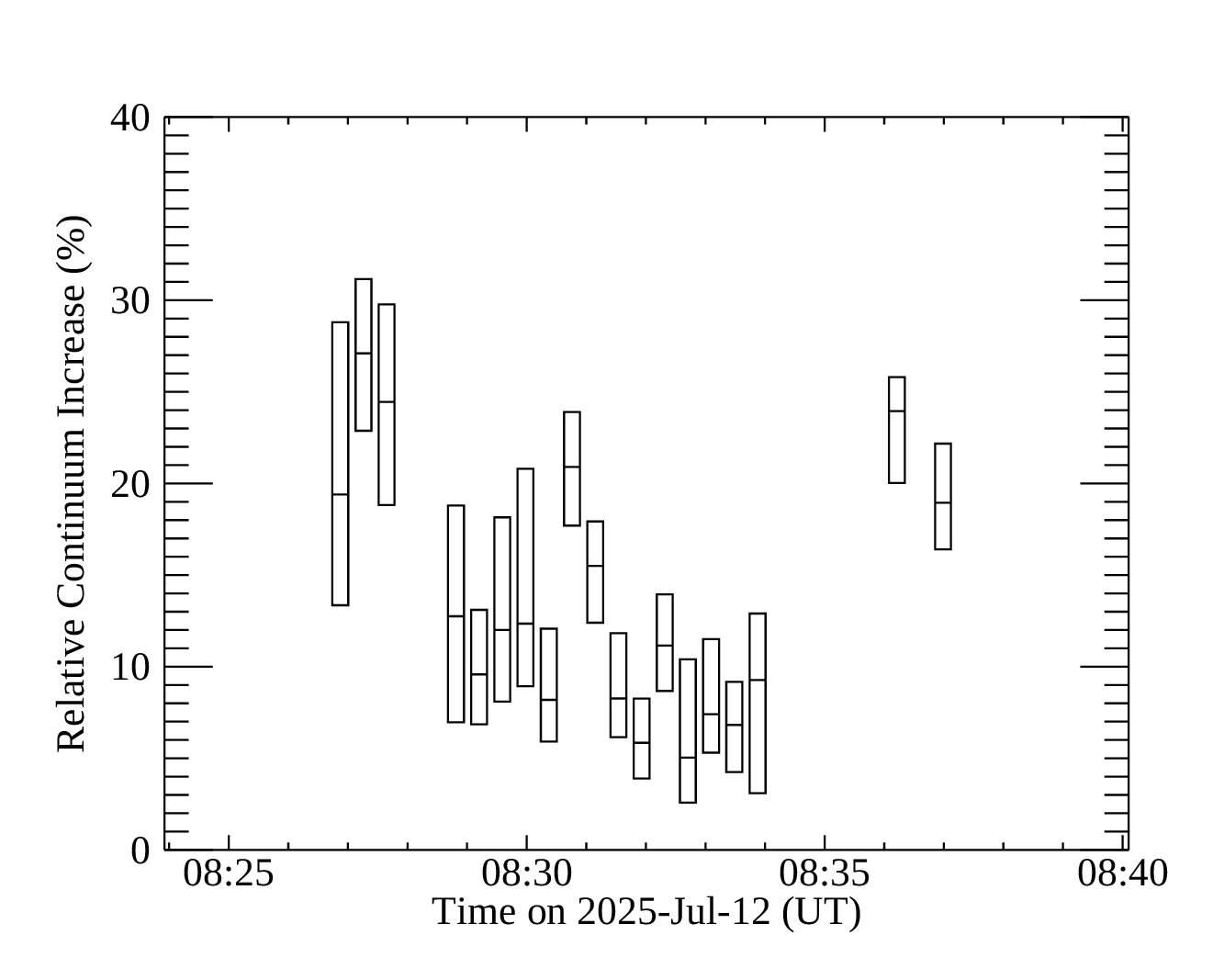}
    \caption{Statistical characteristics of the WL excess intensity (in the \ion{K}{I} continuum) in flare~2 as a function of time (following the same style as Fig.~\ref{fig:wl_intensity_box_1}). The whiskers have been removed for more clarity.}
    \label{fig:wl_intensity_box_2}
\end{figure}

\begin{figure}
    \centering
    \includegraphics[width=0.49\textwidth]{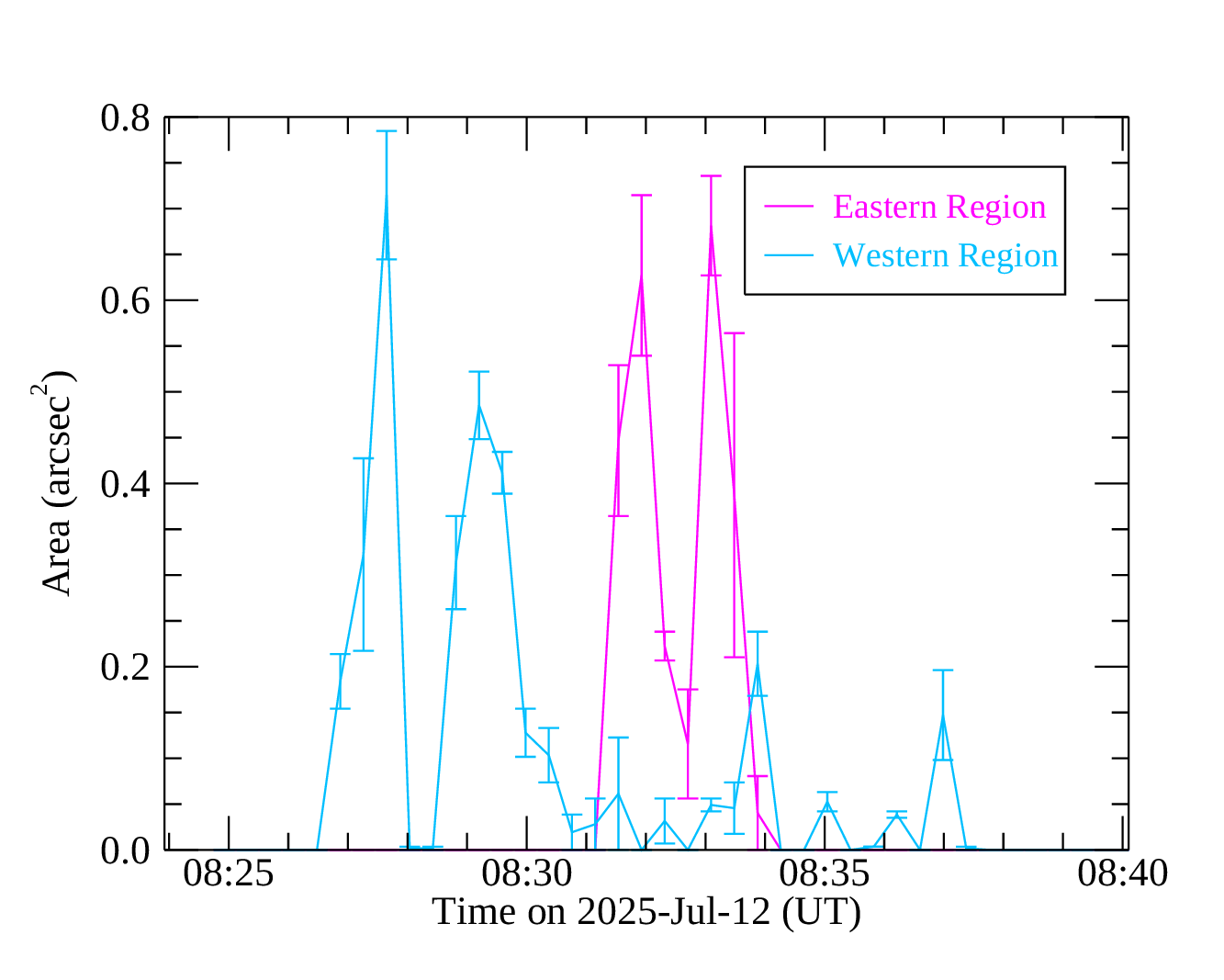}
    \caption{Temporal evolution of the WL area (in the \ion{K}{I} continuum) for flare~2 (following the same style as Fig.~\ref{fig:wl_area_1}).}
    \label{fig:wl_area_2}
\end{figure}

\begin{figure}
    \centering
    \includegraphics[width=0.49\textwidth]{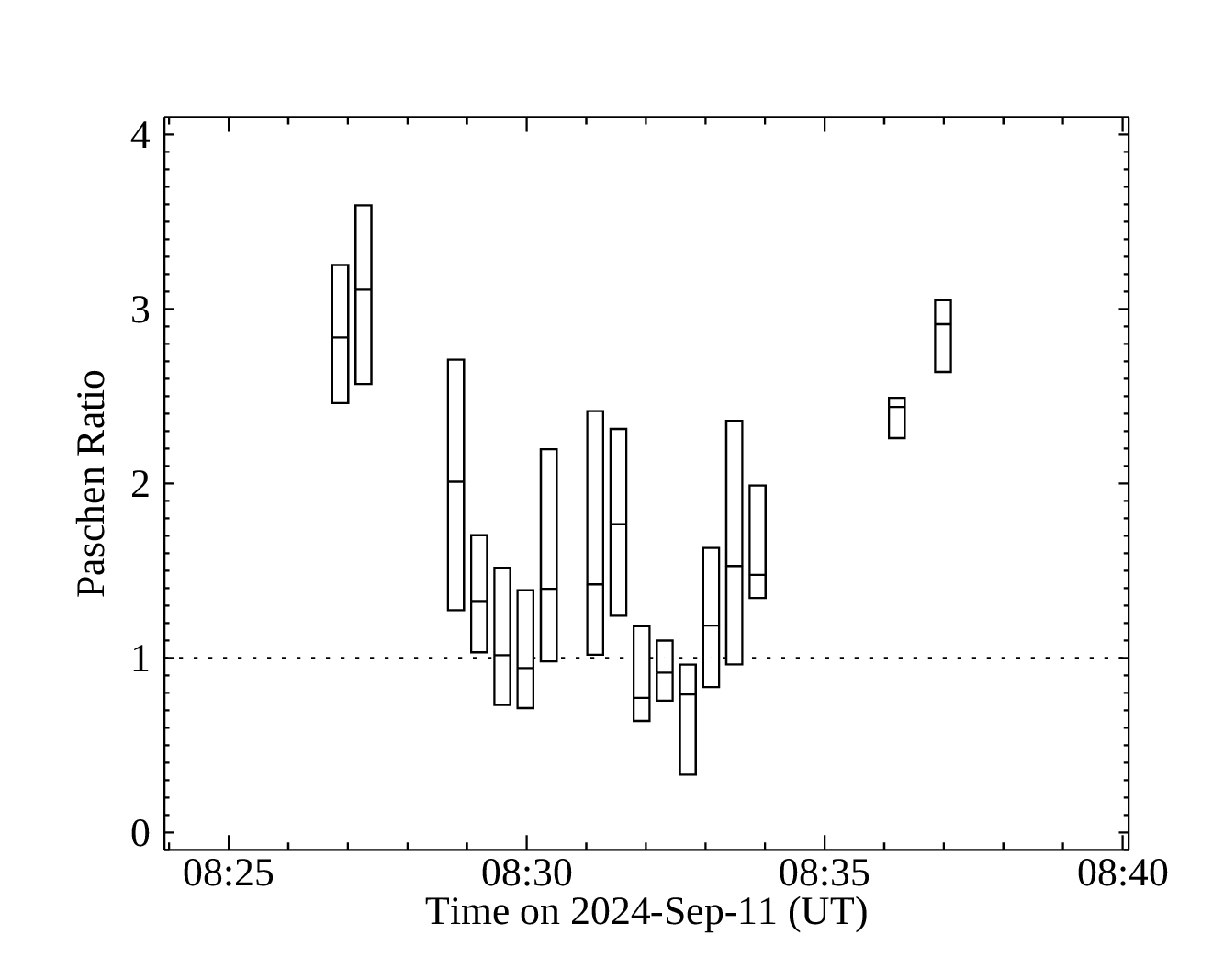}
    \caption{Distribution of the Paschen ratio across WL pixels in flare~2 (following the same style as Fig.~\ref{fig:paschen_ratio_1}). The whiskers have been removed for more clarity.}
    \label{fig:paschen_ratio_2}
\end{figure}

Unlike in flare~1, we captured parts of the onset phase for this flare, which happened in the wake of the declining phase of a C4.8~flare in the same region. Continuum signatures appear in the continua around all three spectral lines. Figure~\ref{fig:flare_general_cwr_2} confirms that the FOV captures both distinct kernels. Similar to flare~1, we calculated subregion light curves of the two locations (ER and WR) of WL enhancements, as displayed in Fig.~\ref{fig:goes_sst_2}. In the bottom two panels, we have marked the two distinct phases of WL enhancements (based on our inspection of the pixel light curves), lasting about 7~min and 2~min, respectively. WL emission initiates prior to the impulsive rise of the M-class SXR flux. This timing anomaly likely stems from the superposition of the ongoing C-class decay phase and the M-class onset. The later half of the early WL window aligns with increased HXR emissions (as derived from the SXR light curve). The increases during the second window of WL enhancements trail the HXR increases by a few minutes. During the first episode of WL enhancements, the maximum WL increase in the ER (compared to a two-minute window at the start of the SST observations) is 3.4\% (\ion{K}{I}~+1150~m\AA), 7.1\% (\ion{Ca}{II}~+4500 m\AA), and \mbox{6.5\% (\ion{Fe}{I}~+680~m\AA)}. In the WR, the respective values are 4.3\%, 4.5\%, and 7.6\%. The maximum increases for the second WL window are negligible in these subregion light curves.\\
\indent The WL pixel map overlaid onto the \ion{K}{I} continuum image in Fig.~\ref{fig:wl_pixels_2} shows that the WL enhancements (in the \ion{K}{I} continuum) in the ER are weaker than in the WR. Overall, they are comparable in strength to the increases apparent in flare~1, ranging from a few percent up to 44\% (the 25th-75th percentile are shown in Fig.~\ref{fig:wl_intensity_box_2}). The corresponding figures for the \ion{Ca}{II} pseudo-continuum and the \ion{Fe}{I} continuum (included in the appendix) show smaller values for \ion{Ca}{II}, but overall larger values for \ion{Fe}{I}. The enhancements before 08:30~UT are generally larger than those after. We note that we calculated the pixel-wise increases with respect to a 1.5~min window one minute prior to the WL increase in each pixel in order for the general trend of the light curve to be negligible, and the increases may therefore differ from those seen in the subregion lightcurves shown in Fig.~\ref{fig:goes_sst_2} (especially due to the averaging over the whole subregion in the latter).\\
\indent The maximum WL area (per definition in the \ion{K}{I} continuum), as depicted in Fig.~\ref{fig:wl_area_2}, is \mbox{$0.68 \pm 0.06$~arcsec$^2$} in the ER and $0.71 \pm 0.07$~arcsec$^2$ in the WR. The total WL areas are $2.52 \pm 0.52$~arcsec$^2$ and \mbox{$3.35 \pm 0.65$~arcsec$^2$}, respectively. As is depicted in Fig.~\ref{fig:paschen_ratio_2}, the mean Paschen ratio is larger than one for most times, and exceeds two for the early and late stages of WL enhancements.\\
\indent Due to the location of the flare close to the solar limb, the CWR in the region is generally higher than in flare~1. The WL area, however, shows a significant increase in CWR during the WL emission enhancements, as can be seen in the right column of Fig.~\ref{fig:flare_general_cwr_2}.\\


\section{Discussion}


\subsection{Energy transport and timing}

The fact that the WL enhancements in flare~1 are co-temporal with the HXR enhancements and that the flare is compatible with the Neupert effect suggests that both stem from nonthermal electrons. For the umbral region, this is corroborated by the reconstructed HXR images, where the enhancements align well with the WL region. In the pore region, on the other hand, no significant HXR emissions are observed, and mechanisms other than electron beams may play the dominant role in transporting the energy to the lower atmosphere. The weak WL enhancements, which are in fact only detectable in the dark regions of the sunspot (and pore), suggest that the beam flux (or energy flux of other energy transport mechanisms) was insufficient to outshine the bright photospheric granulation. This sets an upper limit on the energy flux reaching the layers responsible for the WL enhancements. We note here that we are not able to comment on the role of accelerated protons or Alfvén waves. For flare~2, we hypothesize that the fact that there are WL emission enhancements before significant changes of the derivative of the SXR occur may be related to the C-class flare happening just a few minutes before the flare under investigation. That flare has most likely caused a change in the temperature and ionization structure of the atmosphere that our flare of interest then injects energy into. In a particle-beam scenario (electron or protons), heated and ionized plasma is easier for the beam particles to penetrate, causing the beam to be able to reach lower layers of the atmosphere and deposit its energy there, without influencing the upper layers too much \citep[cf.][]{2018ApJ...862...76P}. The final phase of WL enhancements in this flare follows a few minutes after the maximum of the derivative of the SXR. We therefore cannot rule out additional heating mechanisms other than electron-beam heating.\\
\indent We have shown that the WL enhancements in both flares are not uniform in space, but show clear kernels with significantly larger WL intensity increases. Temporally speaking, higher enhancements in flare~1 occur near the maximum of the 20--50~keV HXR emissions, again suggesting that the WL increases in this flare are electron-beam related.


\subsection{Spectral line diagnostics}

The significantly larger increases of intensity around the \ion{Ca}{II}~line in flare~1 and parts of flare~2 can be due to several factors, which we lay out in the following. Firstly, the reason could be influences from the line affecting the continuum intensity values. This is corroborated by transient hot chromospheric features showing up in \ion{Ca}{II}~pseudo-continuum images that are not present in the continua around \ion{K}{I} and \ion{Fe}{I}. Secondly, the \ion{Ca}{II}~pseudo-continuum may be formed higher up in the atmosphere, where the flare energy deposition is larger. Finally, there could be cool material blocking the LOS of the \ion{K}{I} and \ion{Fe}{I} continua, but this is unlikely. In flare~2, the \ion{Ca}{II}~pseudo-continuum increases are less pronounced (compared to \ion{K}{I}), which we suggest to be due to optically thick material blocking the LOS. Weaker enhancements and/or the different LOS in this flare (which results in plasma velocities up- or downward in the atmosphere not influencing our measurements too much) present alternative explanations.\\
\indent For most WL pixels in flare~1, the \ion{K}{I} and \ion{Fe}{I} lines stay in modestly strong absorption, suggesting weak heating of the upper photospheric layers. We note, however, that many pixels show shallow profiles at the WL excess maximum for both lines, so strong photospheric heating is present in these cases. Generally, the \ion{K}{I}~profiles are shallower than those of \ion{Fe}{I}, which is in line with the \ion{K}{I}~line forming higher up in the atmosphere \citep[this has so far only been shown for quiet-Sun cases, see e.g., \citealt{2017MNRAS.470.1453Q}, but for flares the formation height of \ion{Fe}{I} changes, see][]{2018ApJ...857L...2H}. Strong heating is also apparent at the formation height of the \ion{Ca}{II} line. Here, many profiles show redshifted central emission or an unshifted core with a stronger red-wing emission compared to the blue. According to the response functions presented by \cite{2021A&A...649A.106Y} and \cite{2025A&A...699A.121K}, these characteristics are indicative of mass motions between $\log \tau \approx -2$ and $\log \tau \approx -4$, i.e., in the lower chromosphere. We therefore interpret these features as signatures of chromospheric condensation. The profiles with blueshifted emission peaks can then be qualitatively explained by chromospheric evaporation at higher layers than the condensation \citep[again following the response functions of][]{2021A&A...649A.106Y,2025A&A...699A.121K}, while profiles that show two distinct peaks in the blue and red wing get contributions from both features. We note here that our observed \ion{Ca}{II} profiles generally show a larger central emission intensity and stronger Doppler shifts than in \cite{2021A&A...649A.106Y} and \cite{2025A&A...699A.121K}. The fact that many \ion{K}{I} and \ion{Fe}{I} profiles appear redshifted or have a blue excess (higher blue wing than red wing) can then be explained by (i) downward-moving material in the form of a chromospheric condensation, or (ii) upward-moving material from chromospheric evaporation. Chromospheric contributions to the line formation have already been shown by \cite{2021ApJ...915...16M} \citep[and also appear in the response functions of][]{2021A&A...649A.106Y} for the \ion{Fe}{I}~line. This would also explain the observed large increases in the blue-most spectral point of some profiles of \ion{Fe}{I}. Furthermore, the W-shapes of some \ion{Fe}{I} profiles are similar to those in the bottom-right panel of Fig.~2 of \cite{2018ApJ...857L...2H}, who explain them via mass motions happening between 500 and 800 km above the solar surface. The majority of profiles during the WL maximum in flare~2 are also compatible with chromospheric evaporation and condensation, although the viewing angle complicates the interpretation.


\subsection{Reliability of the Paschen jump diagnostic}

The median Paschen ratio for flare~1 is lower than one throughout most of the period of WL enhancements, which does not conform to the established models for WL enhancements in solar flares. We speculate that this is due to the influence from the \ion{Ca}{II}~line on the pseudo-continuum measurements longward of the Paschen jump, as explained above. The Paschen ratio in flare~2 is generally larger, with the mean exceeding unity for most times. Possible reasons for this have been laid out above.\\
\indent Our results demonstrate that identifying the emission mechanism (photospheric H$^-$ vs. chromospheric recombination) using narrowband filters near broad chromospheric lines comes with difficulties. Future studies require truly line-free continuum windows (e.g., via spectrographs) to unambiguously measure the Paschen jump.


\section{Summary and conclusions}

We present two case studies of WL flares observed with the SST. The results of the timing analysis between WL and HXR enhancements for flare~1 as well as the validity of the Neupert effect for this flare are in line with the WL emission in this event stemming from electron-beam precipitation. In the pore, the misalignment of reconstructed HXR enhancements and WL increases suggests that other energy transport mechanisms might be dominant. The WL enhancements prior to significant changes in the SXR in flare~2 may be related to the C-class flare occurring a few minutes before. The preheated and ionized atmosphere reduces the effective stopping column density, facilitating the penetration of lower-energy electrons into the photospheric layers. The second phase of WL enhancements in this flare may be caused by a heating mechanism other than an electron beam.\\
\indent Our observed WL enhancements range from a few percent to tens of percent. The enhancements are only visible against the darker (pen-) umbral background, and no WL detections in the quiet Sun could be made.\\
\indent A Paschen ratio of less than unity for the majority of WL pixels, as it is the case for flare~1, is inconsistent with hydrogen recombination or H$^-$ emission models, indicating that the \ion{Ca}{II} pseudo-continuum is significantly contaminated by line-wing opacity. Simultaneously, we see large intensity increases for the blue-most spectral point of \ion{Fe}{I}~6173~\AA, which might be the result of an additional chromospheric contribution to the line formation. For flare~2, an occultation of WL enhancements by cool chromospheric material occurred, lowering the measured WL intensities in the \ion{Ca}{II}~pseudo-continuum and raising the observed values of the Paschen ratio.\\
\indent Most of the profiles of \ion{Ca}{II}~8542~\AA, \ion{K}{I}~7699~\AA, and \ion{Fe}{I}~6173~\AA~in the WL regions are consistent with chromospheric condensation in the lower chromosphere and evaporation in the layers above that. The \ion{Fe}{I}~6173~\AA~profiles are generally more complex, possibly due to stronger chromospheric contributions to the line. Nonetheless, our observations show that the continua around the \ion{K}{I}~7699~\AA~and the \ion{Fe}{I}~6173~\AA~line represent continua free of significant line influence, while the pseudo-continuum of the \ion{Ca}{II}~8542~\AA~line is likely to experience a certain degree of alteration from the line itself. We conclude that the \ion{Ca}{II}~8542~\AA~far wing is not a reliable proxy for the continuum in flaring conditions. Unambiguous identification of the emission mechanisms requires strictly line-free continuum observations longward of the Paschen jump.


\begin{acknowledgements}
    This work has been supported by the Research Council of Norway through its Centers of Excellence scheme, project number 262622. The Swedish 1-m Solar Telescope (SST) is operated on the island of La Palma by the institute for Solar Physics of Stockholm University in the Spanish Observatorio del Roque de los Muchachos of the Instituto de Astrofísica de Canarias. The SST is co-funded by the Swedish Research Council as a national research infrastructure (registration number 4.3-2021-00169). The ASO-S mission is supported by the Strategic Priority Research Program on Space Science of the Chinese Academy of Sciences, Grant No. XDA15320000. The authors would also like to thank Dr. Yang Su for their help with ASO-S HXI data, in particular imaging.
\end{acknowledgements}


\bibliographystyle{aa}
\bibliography{paper_bib}

\begin{appendix}

\section{CRISP transmission profiles}

\begin{figure}[ht]
    \includegraphics[width=0.49\textwidth,trim={0 0.8cm 0 2cm},clip]{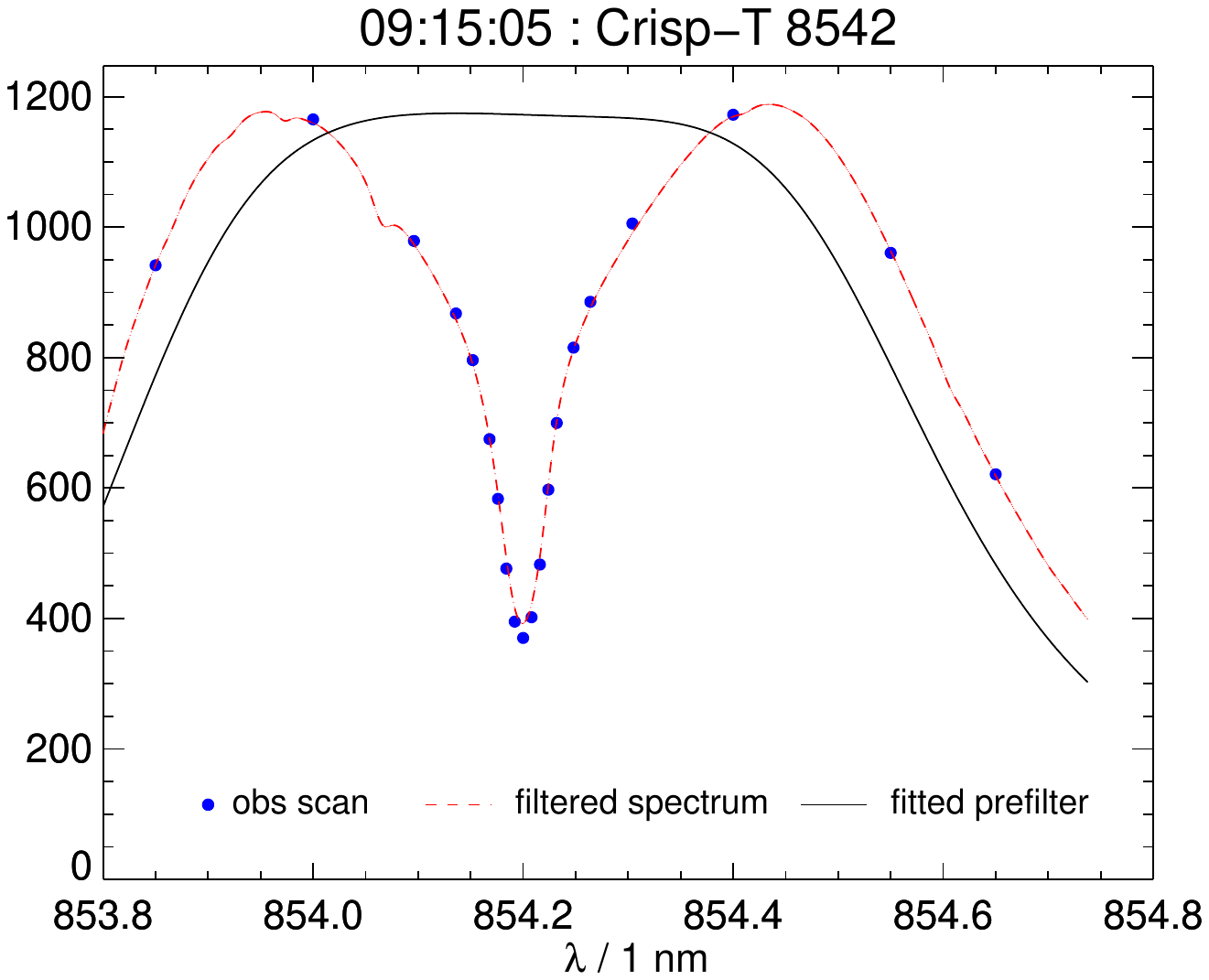}
    \caption{CRISP-T 8542~\AA~prefilter transmission profile in arbitrary units. The spectral points used for flare~2 in this work are shown as blue dots.}
    \label{fig:prefilter}
\end{figure}

In Fig.~\ref{fig:prefilter}, we show the transmission profile of the CRISP-T 8542~\AA~prefilter. Our chosen spectral points used for flare~2 are marked along the curve. This measurement of the prefilter transmission profile was part of the SSTRED \citep{2021A&A...653A..68L} data processing and was done on a reference disk center observation acquired shortly after the observation of flare~2. 

\section{Seeing during the observations}\label{sec:seeing}

\begin{figure}[ht]
    \includegraphics[width=0.47\textwidth,trim={0 0.8cm 0 1.5cm},clip]{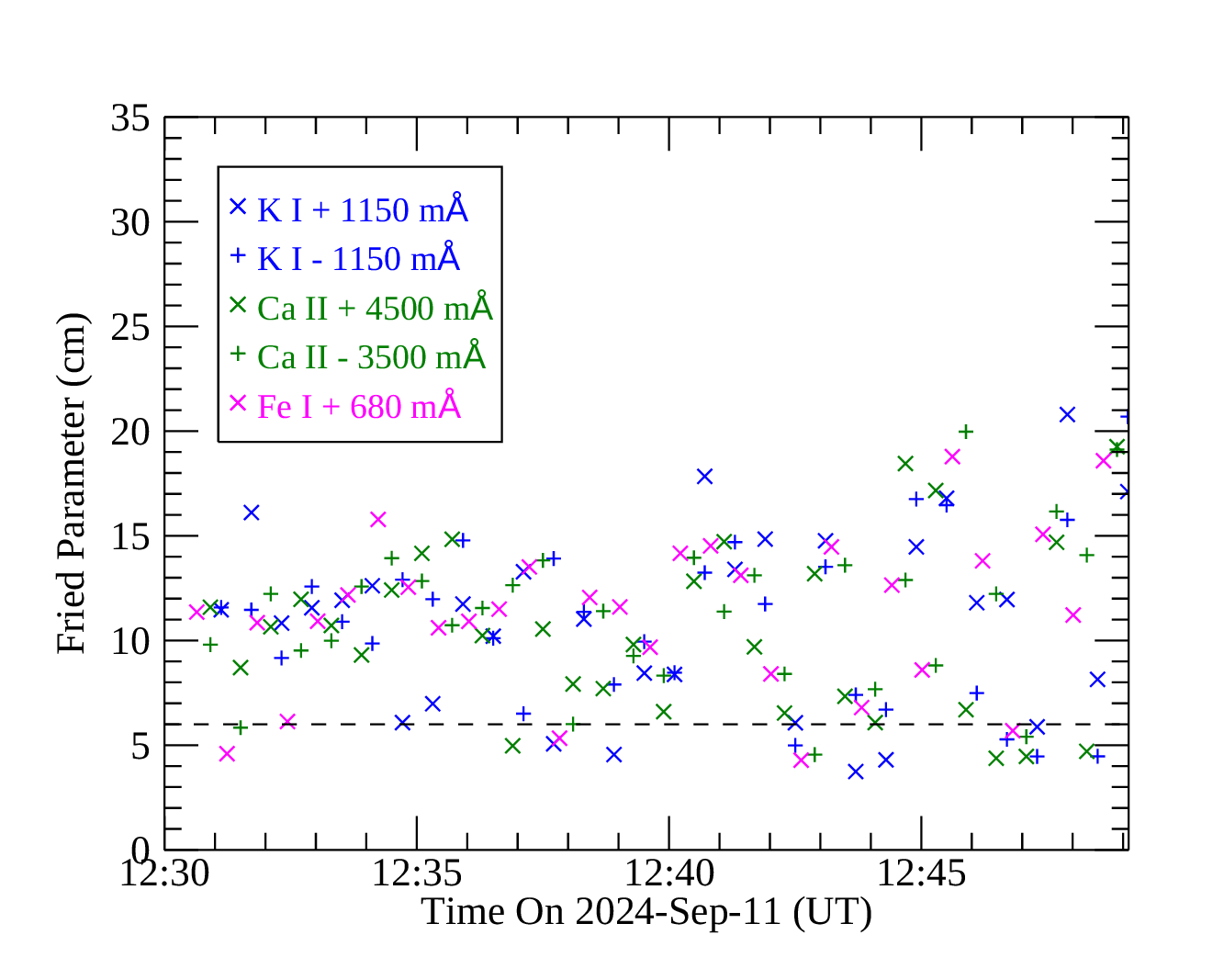}
    \caption{Fried parameter as a function of time for the (pseudo-) continuum around the \ion{K}{I}~7699~\AA~(blue), the \ion{Ca}{II}~8542~\AA~(green), and the \ion{Fe}{I}~6173~\AA~(magenta) line. The red (pseudo-) continuum points are marked with a crosses, while pluses signify the blue (pseudo-) continuum. The horizontal dashed black line marks the level below which observations have been rejected throughout this work.}
    \label{fig:fried_1}
\end{figure}

\begin{figure}[ht]
    \includegraphics[width=0.47\textwidth,trim={0 0.8cm 0 1.5cm},clip]{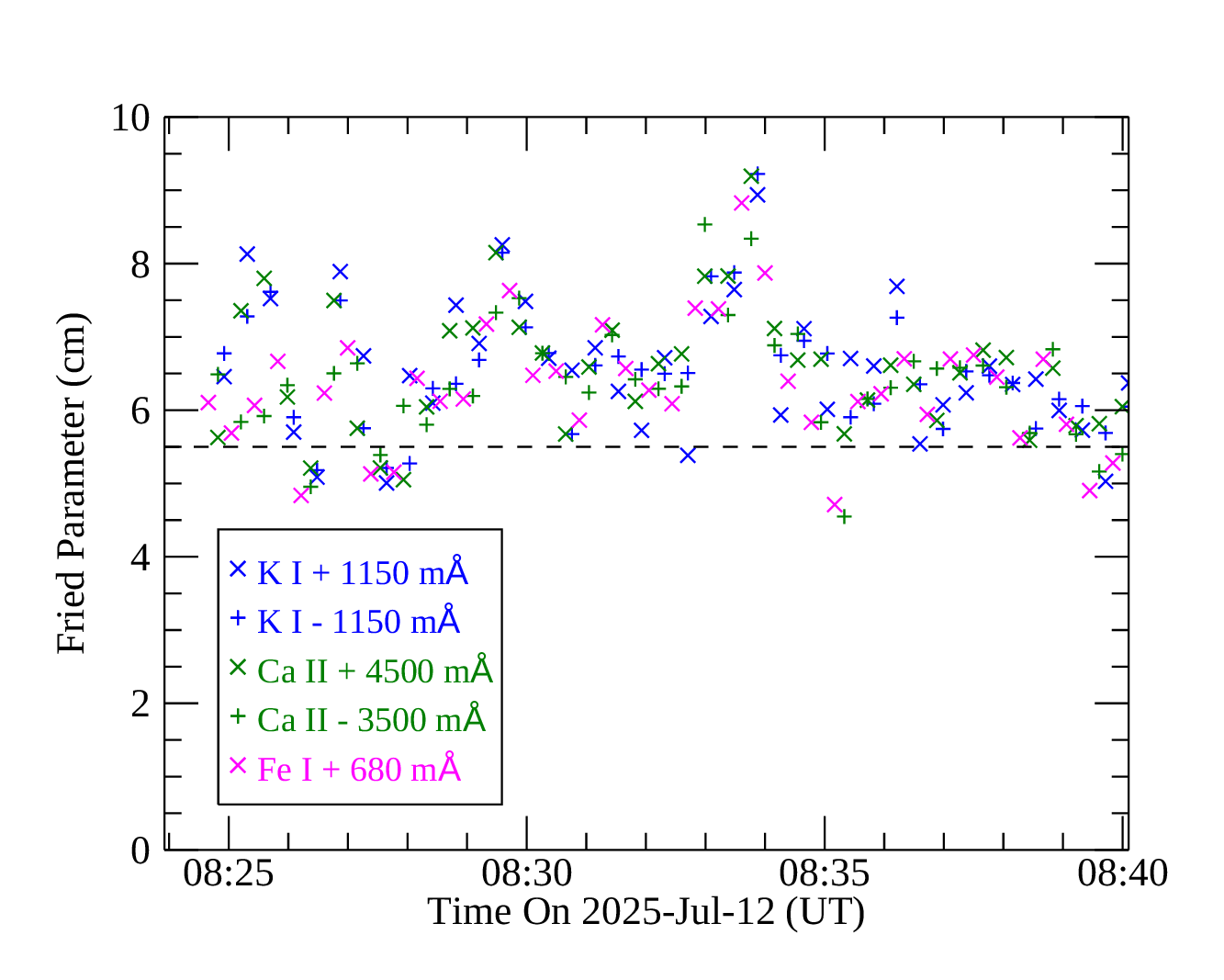}
    \caption{Fried parameter as a function of time for flare~2 (following the same style as Fig.~\ref{fig:fried_1}).}
    \label{fig:fried_2}
\end{figure}

In Fig.~\ref{fig:fried_1}, we show the Fried parameter, $r_0$, as a function of time during the observation of flare~1. We discard observations with $r_0 < 6$~cm (marked by the dashed black line). The same is shown for flare~2 in Fig.~\ref{fig:fried_2} with a threshold of $r_0 < 5.5$~cm. The Fried parameter is measured by the SST adaptive optics system \citep{2024A&A...685A..32S}. 


\section{Spectral profiles}

\begin{figure}[ht]
    {\includegraphics[width=0.49\textwidth]{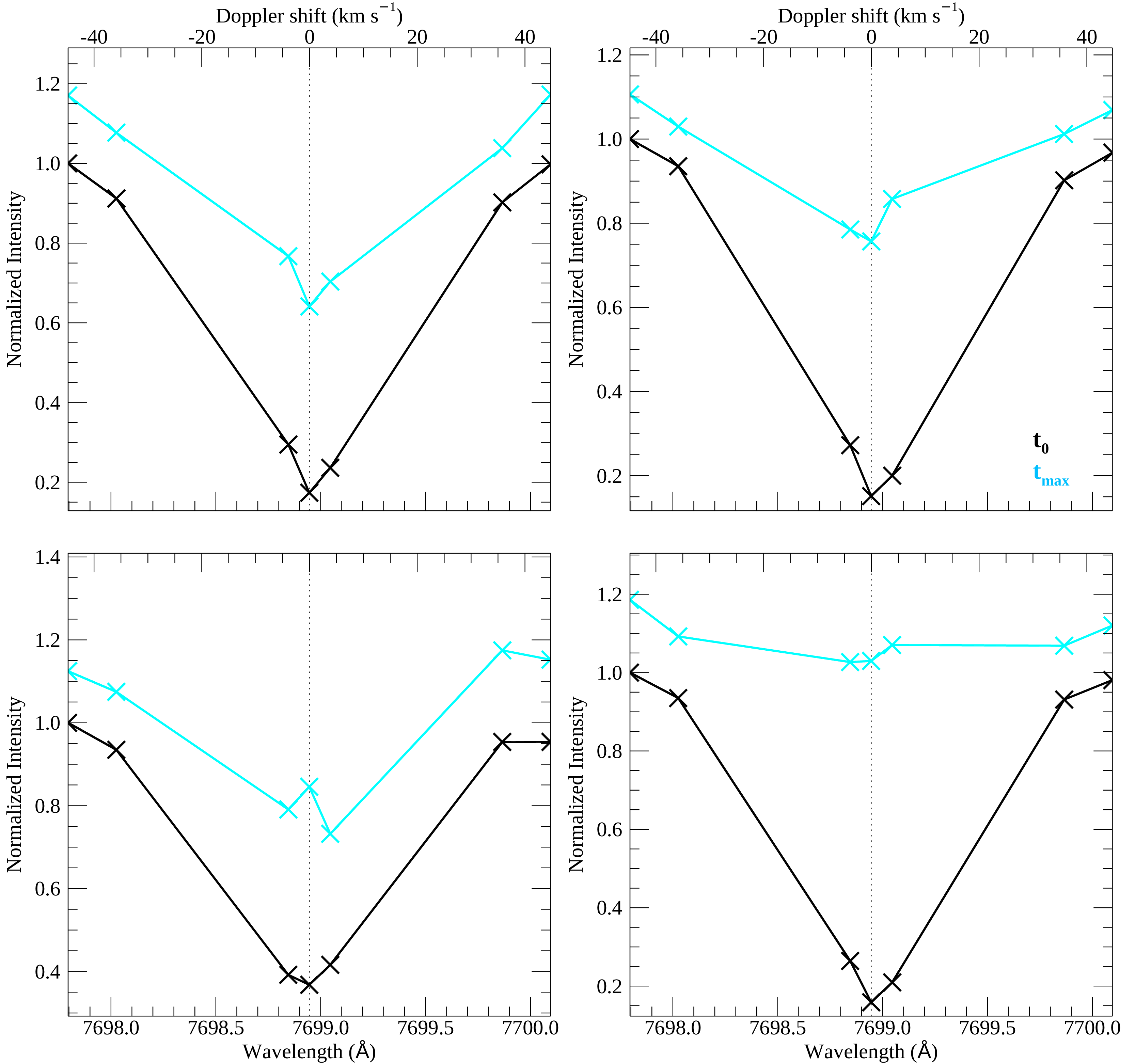}}
    \caption{Example spectral profiles of the \ion{K}{I}~7699~\AA~line for flare~1. In all panels, black profiles correspond to the start time of our observations, whereas cyan profiles were taken at the time of maximum WL enhancements. Crosses mark the measured wavelength points. The vertical dotted line in each panel marks the location of the unshifted line core.}
    \label{fig:profiles_ki}
\end{figure}

\begin{figure*}
    \resizebox{\hsize}{!}
    {\hspace{2.4cm}\includegraphics[width=0.99\textwidth]{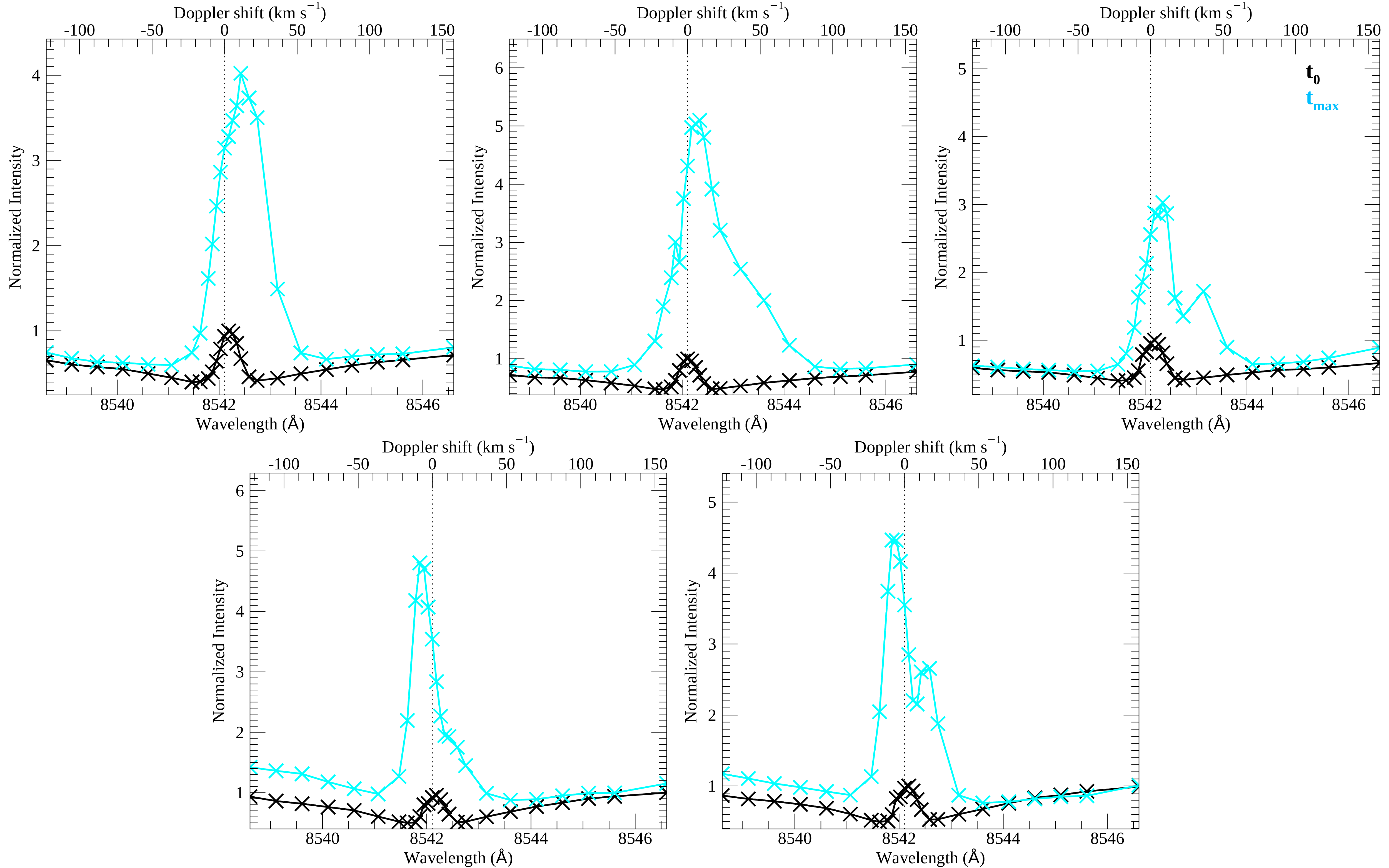}\hspace{2.4cm}}
    \caption{Example spectral profiles of the \ion{Ca}{II}~8542~\AA~line for flare~1. In all panels, black profiles correspond to the start time of our observations, whereas cyan profiles were taken at the time of maximum WL enhancements. Crosses mark the measured wavelength points. The vertical dotted line in each panel marks the location of the unshifted line core.}
    \label{fig:profiles_caii}
\end{figure*}

\begin{figure*}
    \resizebox{\hsize}{!}
    {\hspace{2.4cm}\includegraphics[width=0.99\textwidth]{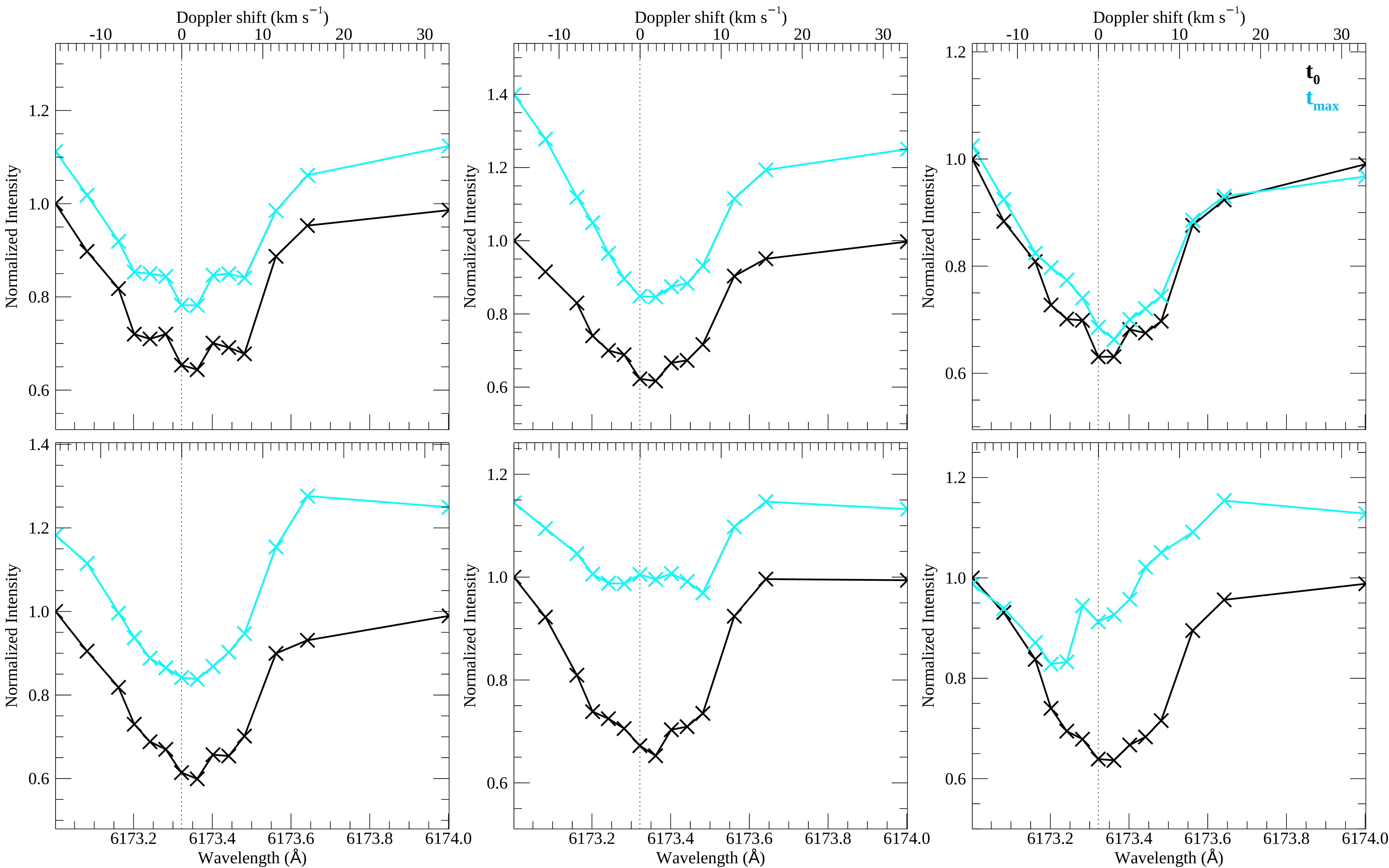}\hspace{2.4cm}}
    \caption{Example spectral profiles of the \ion{Fe}{I}~6173~\AA~line for flare~1. In all panels, black profiles correspond to the start time of our observations, whereas cyan profiles were taken at the time of maximum WL enhancements. Crosses mark the measured wavelength points. The vertical dotted line in each panel marks the location of the unshifted line core.}
    \label{fig:profiles_fei}
\end{figure*}

Figures~\ref{fig:profiles_ki}-\ref{fig:profiles_fei} show example profiles of the \ion{K}{I}~7699~\AA~line, the \ion{Ca}{II}~8542~\AA~line, and the \ion{Fe}{I}~6173~\AA~line during our observations, contrasting the pre-flare (black) and flare profiles (cyan).

\newpage

\section{Continuum enhancements in the Fe~I continuum and the Ca~II pseudo-continuum}

\begin{figure*}[ht]
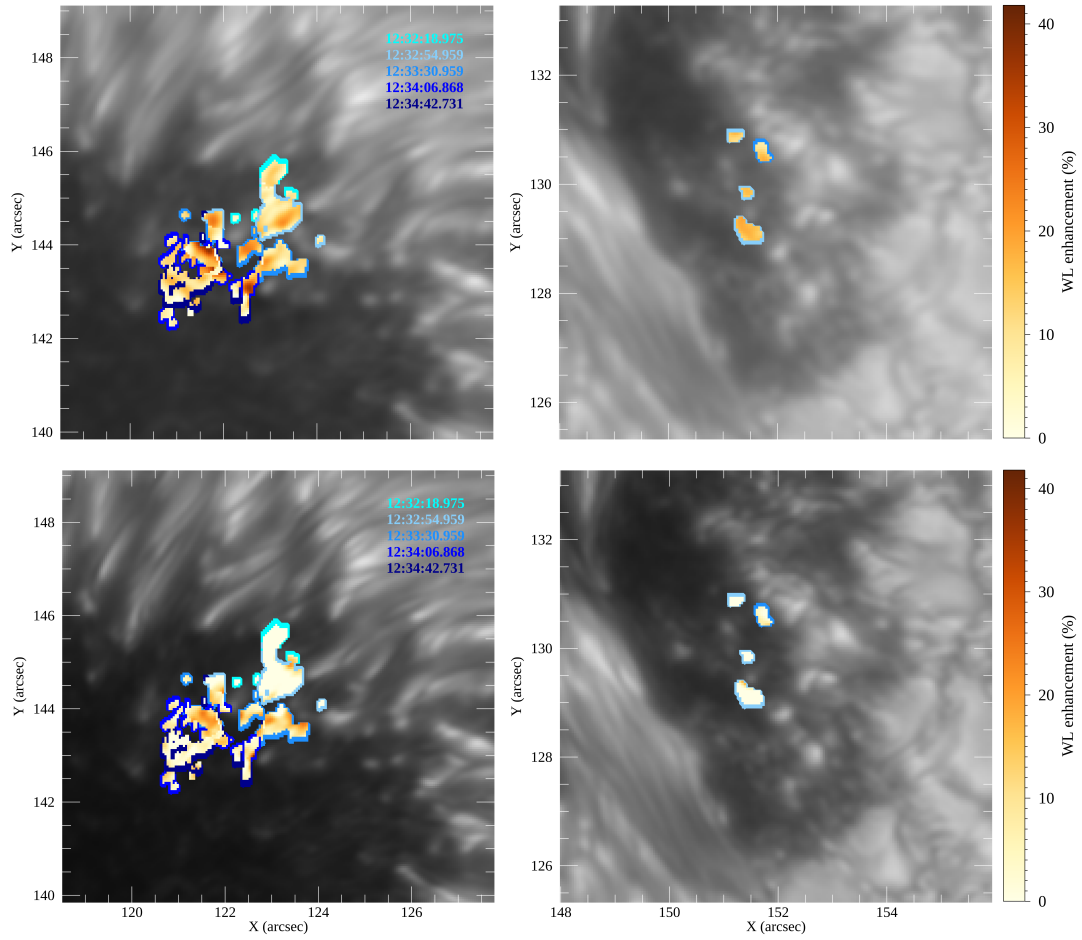

    \resizebox{\hsize}{!}
    {\hspace{2.5cm}\includegraphics[width=0.99\textwidth,trim={0 2.0cm 0 0},clip]{figures/wl_pixels_all_time_intensity_ca_1.png}\hspace{2.7cm}}
    {\hspace*{1.98cm}\includegraphics[width=0.777\textwidth,trim={0 0 0 1.5cm},clip]{figures/wl_pixels_all_time_intensity_fe_1.png}}
    \caption{\ion{Ca}{II} pseudo-continuum (top panels) and \ion{Fe}{I} continuum (bottom panels) intensity enhancement (in percent) for each identified WL pixel for flare~1 overlaid on images of the \ion{Ca}{II} pseudo-continuum. The left panel shows a subregion of the sunspot, while the right panel depicts a zoom into the western part of the pore. Colored contours around the WL areas mark the time of maximum WL enhancement, as indicated in the top-right corner of the left panel. All values are calculated based on a linear interpolation to the time of the respective observation in the \ion{K}{I} continuum.}
    \label{fig:wl_pixels_ca_fe_1}
\end{figure*}

\begin{figure*}[ht]
    \resizebox{\hsize}{!}
    {\hspace{1.0cm}\includegraphics[width=0.32\textwidth,trim={0 1cm 0 1cm},clip]{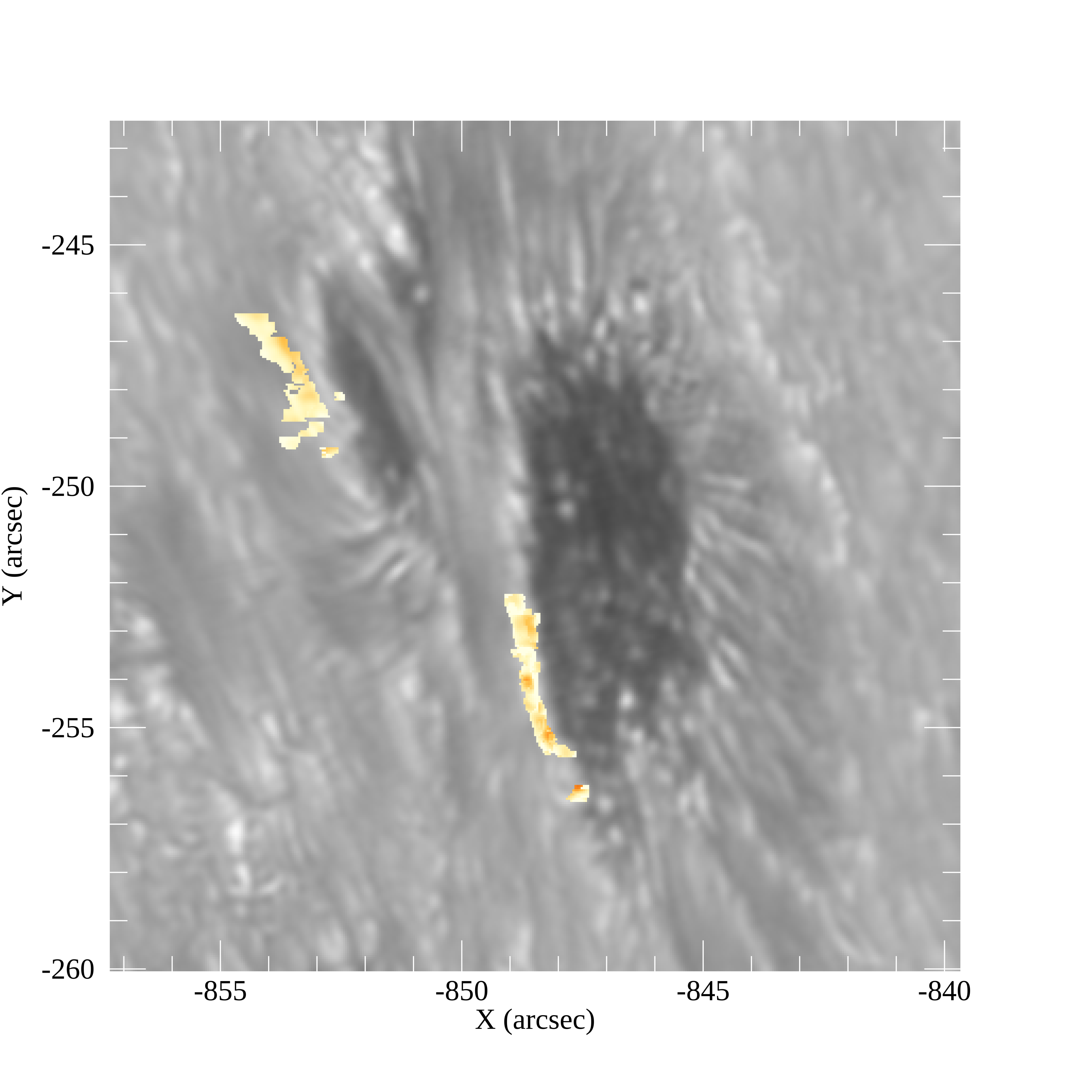}
    \includegraphics[width=0.32\textwidth,trim={0 1cm 0 1cm},clip]{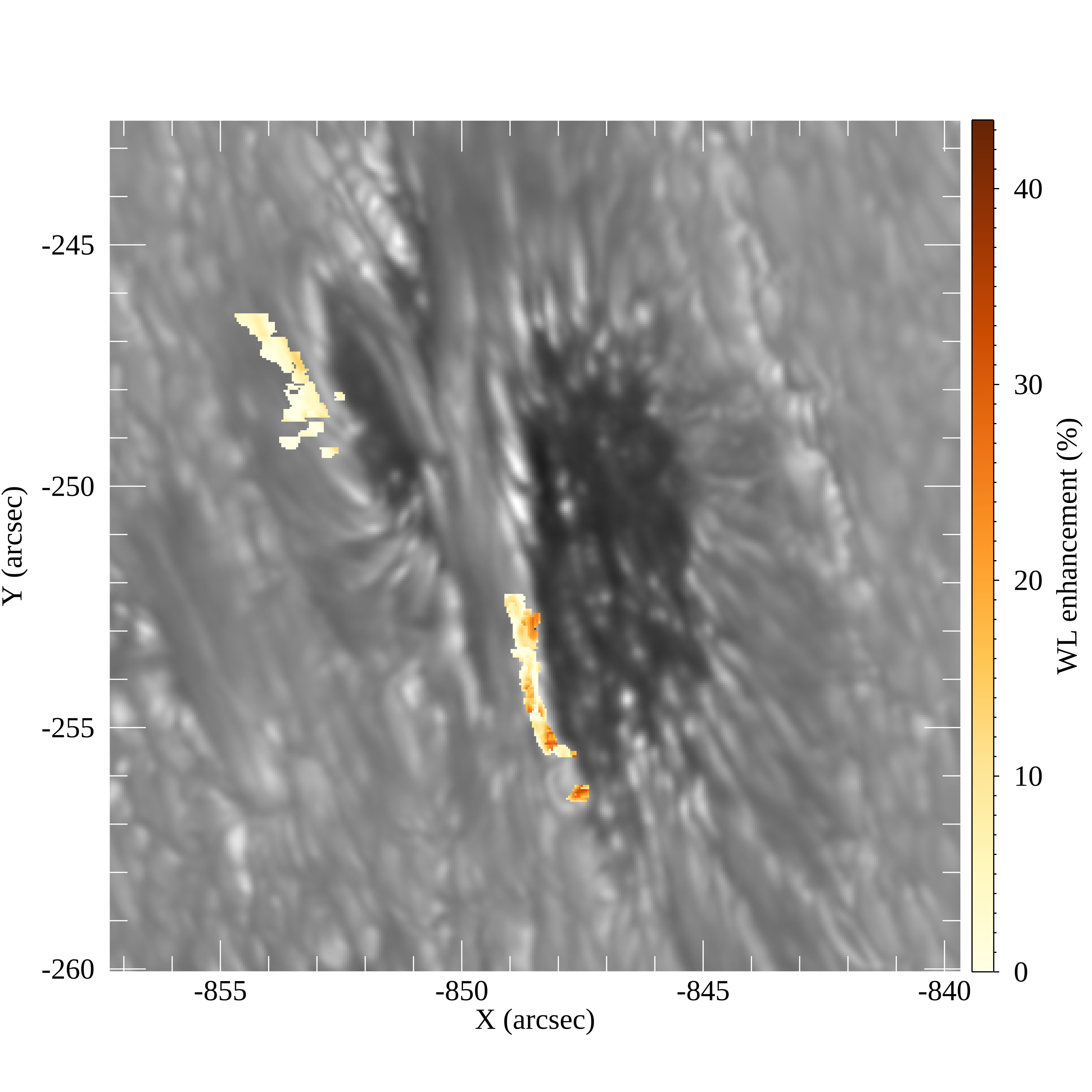}\hspace{1.0cm}}
    \caption{\ion{Ca}{II} pseudo-continuum (left) and \ion{Fe}{I} continuum (right) intensity enhancement (in percent) for each identified WL pixel for flare~2 overlaid on images of the respective continuum. All values are calculated based on a linear interpolation to the time of the respective observation in the \ion{K}{I} continuum.}
    \label{fig:wl_pixels_ca_fe_2}
\end{figure*}

\begin{figure}[ht]
    {\includegraphics[width=0.49\textwidth,trim={0 2cm 0 0},clip]{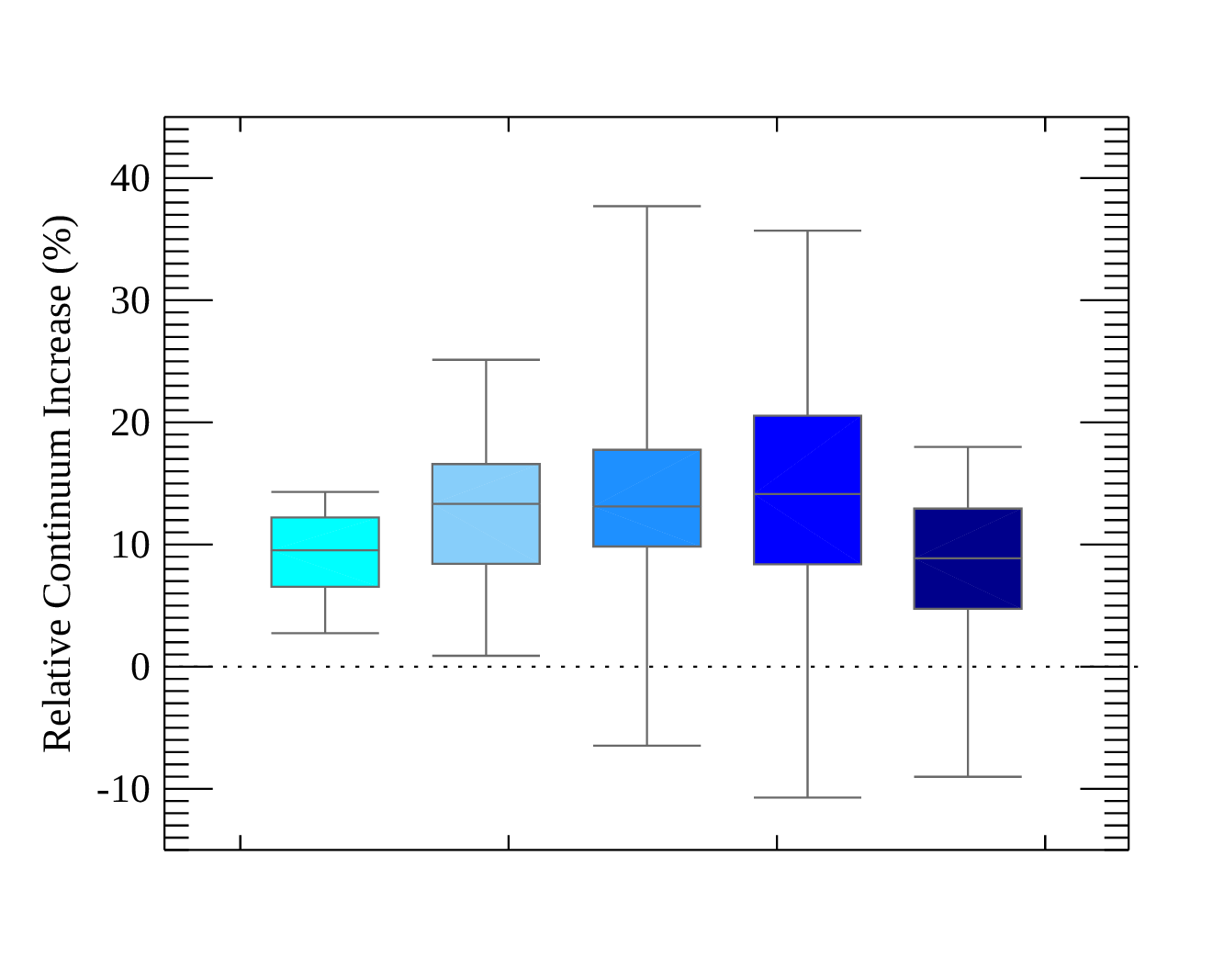}\hspace{0.2cm}}\\
    {\includegraphics[width=0.49\textwidth,trim={0 0 0 2cm},clip]{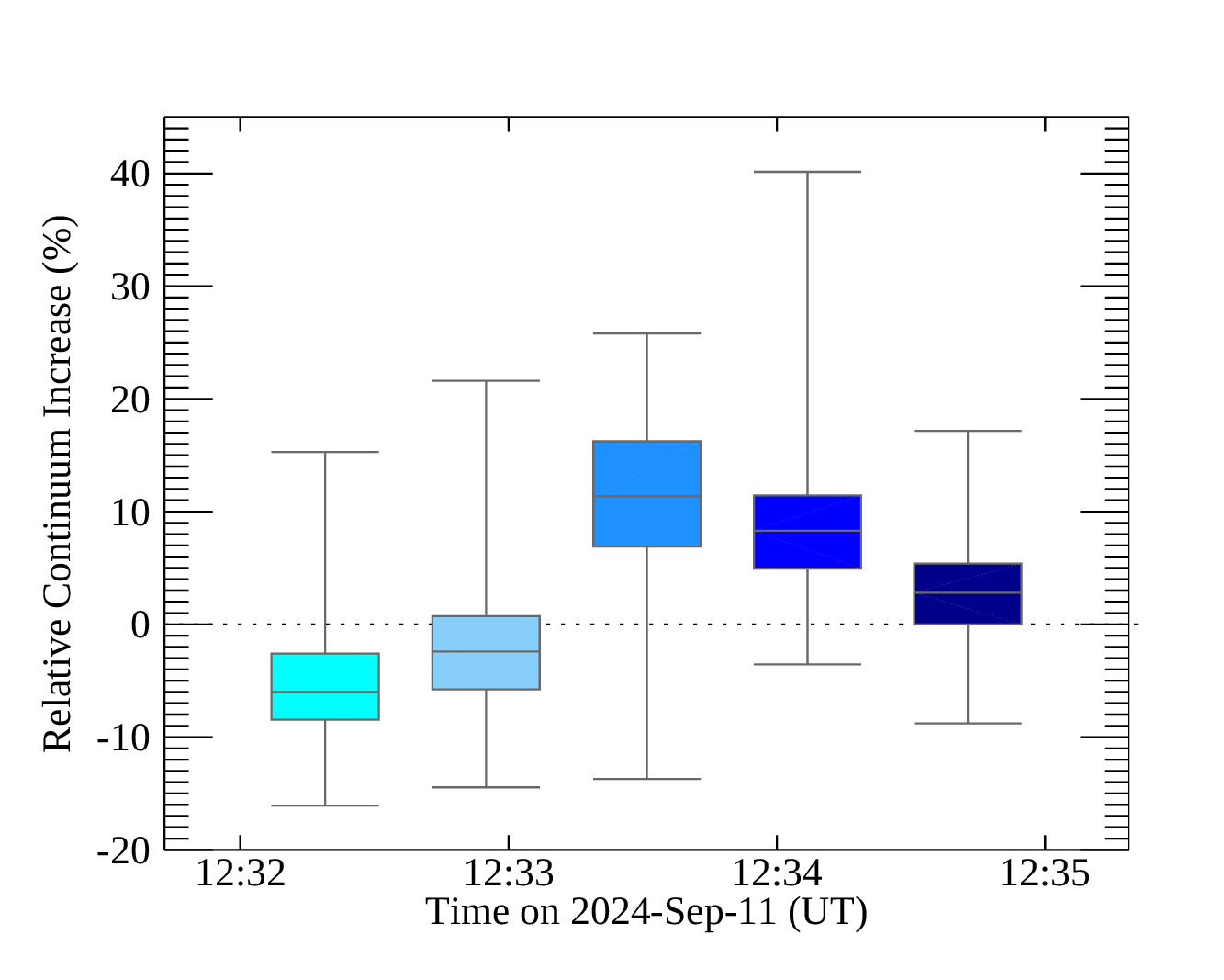}}
    \caption{Statistical characteristics of the WL excess intensity in flare~1 as a function of time (following the same style as Fig.~\ref{fig:wl_intensity_box_1}). Top panel: \ion{Ca}{II} pseudo-continuum. Bottom panel: \ion{Fe}{I} continuum. All values are calculated based on a linear interpolation to the time of the respective observation in the \ion{K}{I} continuum.}
    \label{fig:wl_intensity_box_ca_fe_1}
\end{figure}

\begin{figure}[ht]
    {\includegraphics[width=0.49\textwidth,trim={0 1.7cm 0 0},clip]{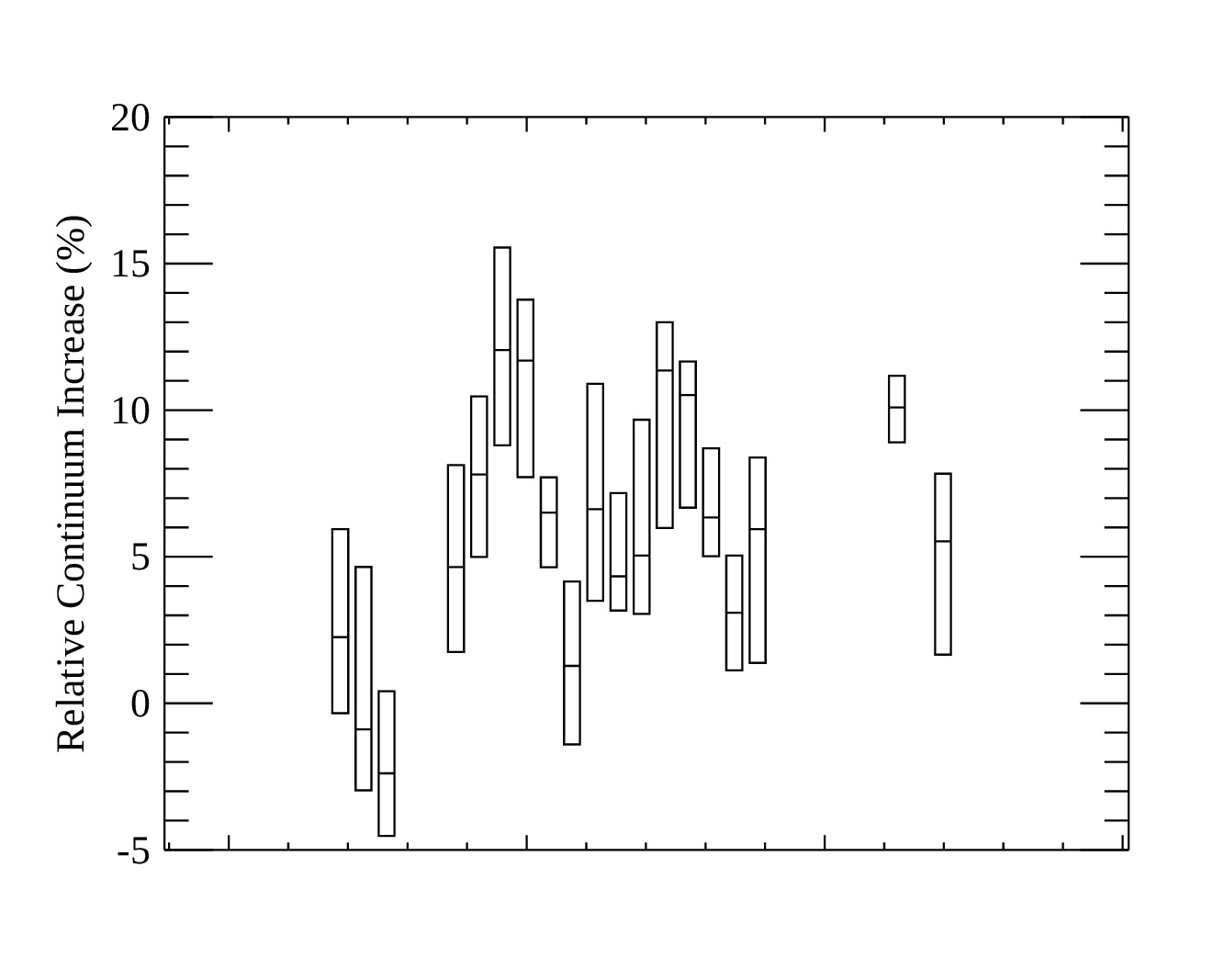}\hspace{0.2cm}}\\
    {\includegraphics[width=0.49\textwidth,trim={0 0 0 1.7cm},clip]{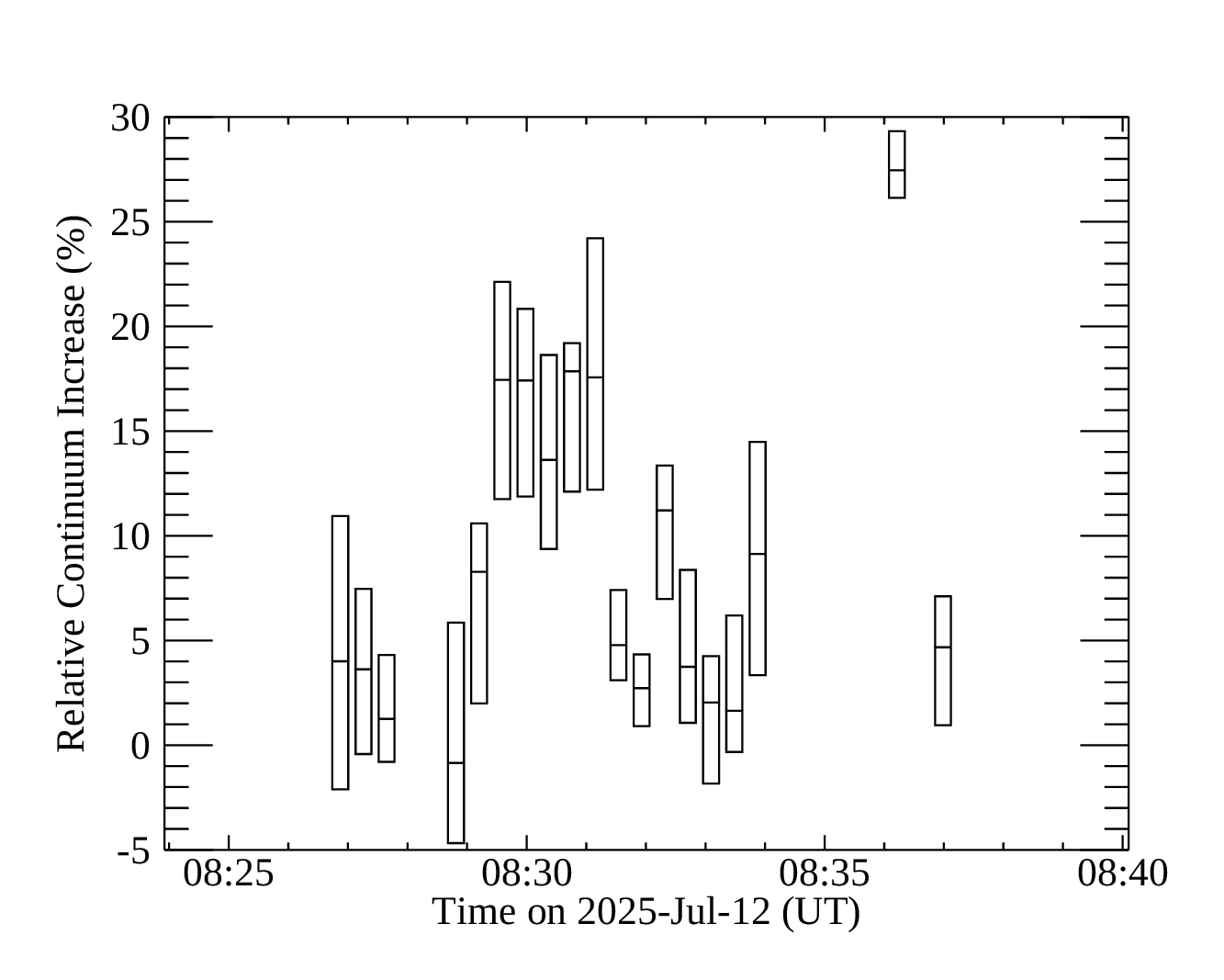}}
    \caption{Statistical characteristics of the WL excess intensity in flare~2 as a function of time (following the same style as Fig.~\ref{fig:wl_intensity_box_2}). Top panel: \ion{Ca}{II} pseudo-continuum. Bottom panel: \ion{Fe}{I} continuum. All values are calculated based on a linear interpolation to the time of the respective observation in the \ion{K}{I} continuum. The whiskers have been removed for more clarity.}
    \label{fig:wl_intensity_box_ca_fe_2}
\end{figure}

Figures~\ref{fig:wl_pixels_ca_fe_1} and \ref{fig:wl_pixels_ca_fe_2} depict the WL intensity enhancement in the \ion{Ca}{II} pseudo-continuum and the \ion{Fe}{I} continuum for flare~1 and flare~2, respectively, overlaid onto corresponding (pseudo-) continuum images. Figures~\ref{fig:wl_intensity_box_ca_fe_1} and \ref{fig:wl_intensity_box_ca_fe_2} show diagrams of the continuum enhancements around both lines for both flares. The enhancements follow generally the same trend, but have varying amplitudes and slightly different locations, possibly owing to the interpolation not perfectly reflecting the real continuum enhancement occurring between measured timesteps.

\end{appendix}

\end{document}